\documentclass[8pt,a4paper]{extarticle}
\usepackage[a4paper]{geometry}
\usepackage[utf8]{inputenc}
\usepackage{graphicx}
\usepackage{csquotes}
\usepackage{hyperref}
\usepackage{authblk}
\usepackage{booktabs}
\usepackage{longtable}
\usepackage{adjustbox}
\usepackage{natbib}
\usepackage{multirow}
\setcitestyle{authoryear}
\usepackage{color}
\usepackage{xcolor}
\usepackage[normalem]{ulem}
\usepackage{etoolbox}
\robustify\itshape
\usepackage{siunitx}
\usepackage{pifont}
\usepackage{makecell}
\hypersetup{pdfborder={0 0 0} } 
\newcommand{\cmark}{\textcolor[HTML]{3969AC}{\ding{51}}}%
\newcommand{\xmark}{\textcolor[HTML]{E73F74}{\ding{55}}}%

\title{Adaptive continuity-preserving simplification of street networks}

\author[1]{Martin Fleischmann}
\author[2,3]{Anastassia Vybornova}
\author[4]{James D. Gaboardi}
\author[1]{Anna Brázdová}
\author[1]{Daniela Dančejová}

\affil[1]{Department of Social Geography and Regional Development, Charles University, Czechia}
\affil[2]{Copenhagen Center for Social Data Science (SODAS), University of Copenhagen, Denmark}
\affil[3]{NEtworks, Data and Society (NERDS), Data Science Section, IT University of Copenhagen, Denmark}
\affil[4]{Geospatial Science and Human Security, Oak Ridge National Laboratory, Oak Ridge, USA}

\date{May 2025}

\begin{document}

\maketitle

\begin{abstract}
Street network data is widely used to study human-based activities and urban structure. Often, these data are geared towards transportation applications, which require highly granular, directed graphs that capture the complex relationships of potential traffic patterns. While this level of network detail is critical for certain fine-grained mobility models, it represents a hindrance for studies concerned with the morphology of the street network. For the latter case, street network simplification -- the process of converting a highly granular input network into its most simple morphological form -- is a necessary, but highly tedious preprocessing step, especially when conducted manually. In this manuscript, we develop and present a novel adaptive algorithm for simplifying street networks that is both fully automated \textit{and} able to mimic results obtained through a manual simplification routine. The algorithm -- available in the \texttt{neatnet} Python package -- outperforms current state-of-the-art procedures when comparing those methods to manually, human-simplified data, while preserving network continuity.
\end{abstract}

\tableofcontents

\section{Introduction}\label{sec:intro}

Street networks as objects of study are an enduring topic of basic and applied research across a broad spectrum of disciplines -- geography \citep{gaboardi_connecting_2020}, urban morphology \citep{marshall_street_2018}, urban economics \citep{yoshimura_spatial_2021}, transportation planning \citep{lovelace_open_2021}, spatial demography \citep{daniel_gis-based_2021}, and operations research \citep{cororan_user-centric_2025}, to name just a few. In recent years, large-scale analyses of street network data have become increasingly feasible thanks to open data initiatives such as OpenStreetMap and Overture Maps \citep{openstreetmap_contributors_openstreetmap_2025, overture_maps_foundation_overture_2025}, and advances in open-source methods for street network analysis such as \texttt{OSMnx} \citep{boeing_osmnx_2017, boeing_modeling_2024}, \texttt{snkit} \citep{russell_snkit_2024}, and \texttt{spaghetti} \citep{gaboardi_spaghetti_2021}. There is, however, an unresolved methodological challenge when it comes to the \textit{simplification} of street networks (see our working definition of the term below, the illustration in Figure~\ref{fig:original-manual}, and Appendix~\ref{appendix:protocol}). Frequently, the available data is a \textit{transportation-based} representation of the street network, where all traffic lanes and pedestrian or cyclist paths are mapped separately (see Figure~\ref{fig:original-manual}a). However, the input format required by many empirical use cases is a \textit{morphological} representation of the street network, where intersections are represented by single nodes and street segments are represented by single edges (see Figure~\ref{fig:original-manual}b). The process of converting a transportation-based street network representation (Figure \ref{fig:original-manual}a) to a morphological one (Figure~\ref{fig:original-manual}b) is what we call \textit{(street network) simplification}. Street network simplification is of particular relevance for urban morphology \citep{fleischmann_shape-based_2024}, be it street network analysis \citep{porta_network_2006, barthelemy_modeling_2008, boeing_using_2022}, or space syntax studies \citep{karimi_space_2018, el_gouj_urban_2022}. In addition, other applications such as drone flight routing \citep{morfin2023u}, navigation \citep{bongiorno_vector-based_2021} or data visualisation \citep{lovelace_route_2024} share these needs. Notably, a simplified network matters not only for analysis of the network itself, but also for studies concerned with the graph faces, i.e., the urban blocks delineated by the network edges \citep{louf_typology_2014}.

The consequences of using an incorrect street network representation for further analysis vary from marginal to critical. Marginal effects are observed when the street network is just one of many inputs of the analysis. For example, in the case of a combined reflection of urban form and function in \cite{arribas-bel_spatial_2022}, the focus on a wider context and the inclusion of additional data mitigates the errors in the network. In contrast, for studies focusing solely on urban form, the consequences can be more pronounced. For example, \cite{araldi2024multi} use transportation networks as a representation of streetscapes which should be conceptualised as morphological. This leads to a duplication of computation (e.g.~the same street represented by a dual carriageway is computed twice) and occasional artifacts breaking the definition of what a streetscape is when two geometries that should be representing the same street are classified as different urban fabrics (see Figure B1 in \cite{araldi2024multi} for an illustration). When using street networks to delineate urban blocks, this requires specific post-processing to remove the geometries that are, in fact, artifacts of the network representation \citep{louf_typology_2014}). Some types of analyses are actually impossible, as the transportation network may not be capable of correctly representing the conceptual network in need. Lastly, in empirical studies on navigation using GPS trajectories, transportation networks complicate map matching (allocating observation data to linear features), and routes on the same street may be allocated to different network edges due to GPS imprecision only, requiring at least some degree of simplification \citep{bongiorno_vector-based_2021}. 

Manual simplification (for example, in a desktop GIS software like QGIS \citep{QGIS_software}) is cognitively feasible, but in practice prohibitively time-consuming. In contrast, existing tools and methods for automated simplification, while already very useful, do not resolve the challenge of implementing a fully automated, entirely reproducible simplification process. In addition, at the time of writing, to our knowledge there are no available methods -- neither proprietary nor open-source -- for automated street network simplification that require \textit{only} the street network as an input parameter, i.e.~that do not rely on geometry attributes for the simplification process or case-specific parameters. Accordingly, recent studies from widely varying fields have identified the lack of an automated street network simplification method as a significant limitation \citep{uhl_towards_2022, ballo_modeling_2023, aiello_urban_2025}. 

Our work addresses this need by introducing a method which is open-source, robust, computationally efficient, fully automated, entirely reproducible, \textit{and} consistently outperforms all other previous simplification methods. We achieve this by identifying simplification subtasks and tackling them separately in an iterative manner, leveraging previous approaches from the literature \citep{boeing_osmnx_2017, tripathy_open-source_2021, fleischmann_shape-based_2024}. We evaluate our method through a comparison to the manually simplified networks and the performance of the state-of-the-art tool. For a sample of seven functional urban areas \citep{schiavina_ghs-fua_2019} listed in Table~\ref{tab:fuas}, we compute a range of relevant metrics for their manually simplified network vs.~for networks simplified by different approaches, and show that our approach generally both outperforms all previous approaches and comes closest to the ``ground truth'' -- manually simplified networks. For virtually any application in research and policy that uses street network data as input, and is not strictly concerned with routing or transportation, the proposed method can be of great use.

\begin{table}[h!]
\centering
\begin{tabular}{lll}
    City             & Country       & Region / Continent \\
\midrule
    Aleppo           & Syria         & Middle East / Asia \\
    Auckland         & New Zealand   & Oceania / Asia \\
    Douala           & Cameroon      & Africa \\
    Liège            & Belgium       & Europe \\
    Bucaramanga      & Colombia      & S. America \\
    Salt Lake City   & USA           & N. America \\
    Wuhan            & China         & Far East / Asia \\
\bottomrule
\end{tabular}
\caption{The seven Functional Urban Areas (FUAs) selected for our analysis. The sample is selected to represent street network paradigms from varied geographical contexts.}
\label{tab:fuas}
\end{table}

The rest of the paper is organized as follows: We define the term ``simplification'' in the context of this paper, and then review already existing street network simplification methods, with a particular focus on those that have been integrated in the method we propose. We then describe our method. In the results section, we compare outcomes of our method to outcomes of other methods to evaluate their respective performance. We end by discussing limitations of our method and potential next steps.

\section{What is simplification?}\label{sec:what-is-simp}

\begin{figure}[h]
    \centering
    \includegraphics[width=0.95\textwidth]{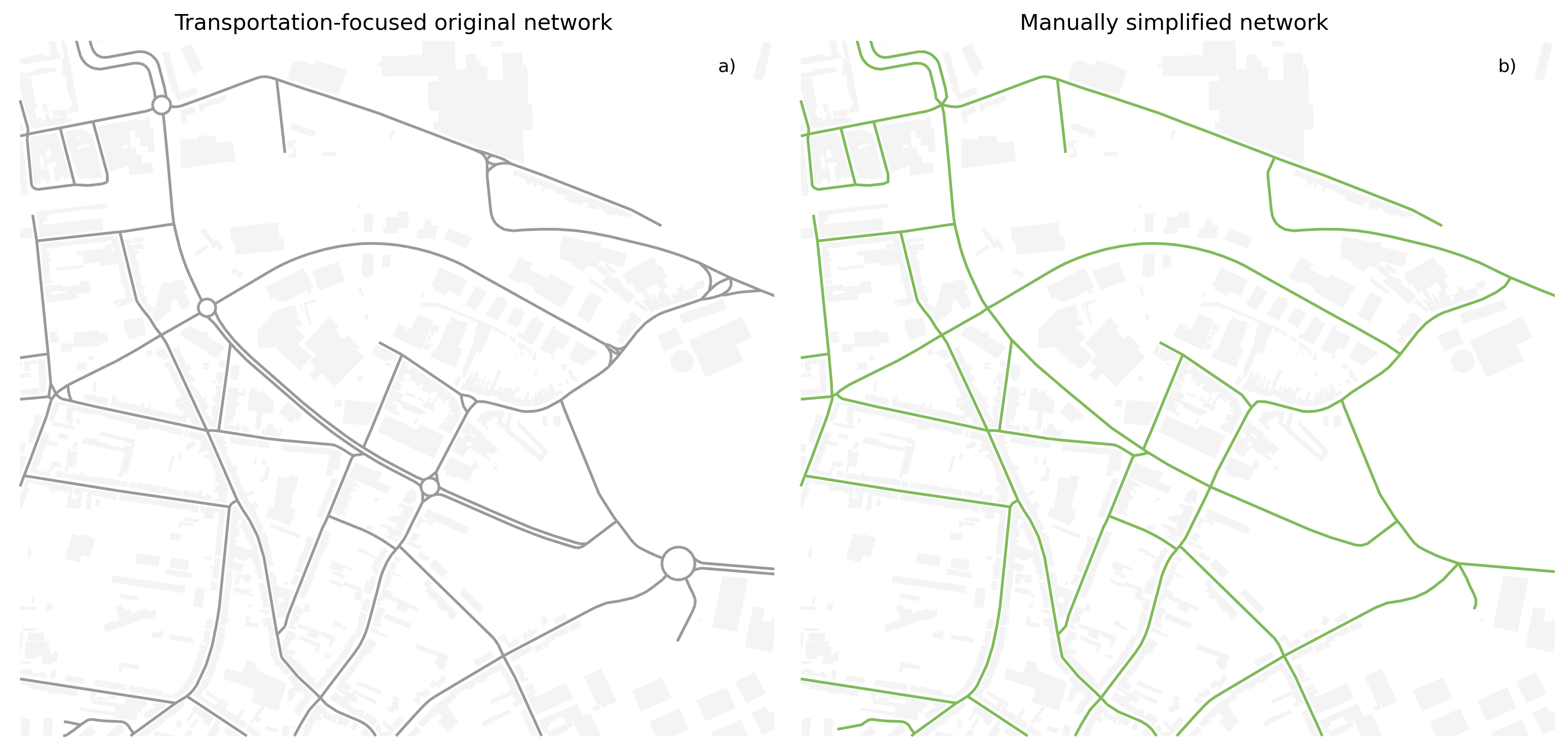}
    \caption{Illustration of the simplification process by the example of a street network fragment in Seraing, Liège (Belgium).~(a)~The left panel shows the input data, downloaded from OpenStreetMap \citep{openstreetmap_contributors_openstreetmap_2025}, before simplification. Due to transportation-focused mapping, the input network contains several intersections, roundabouts, and dual carriageways that are mapped with a high level of granularity.~(b)~The same network after simplification. The level of granularity has been reduced where appropriate, so that every intersection is represented by one node and every street segment is represented by one edge.}
    \label{fig:original-manual}
\end{figure}

Simplification is the process of converting a street network into its most simple morphological representation, where intersections are represented by single nodes and street segments are represented by single edges. Typically, simplification is necessary in cases where a network that has been mapped with a relatively high level of granularity -- as required by transportation-focused applications -- serves as input for an application that requires a lower level of granularity. Simplification includes the removal of parallel edges (dual carriageways), the replacement of detailed transportation-based intersection and interchange mappings by a single node -- or, where this is not possible, by lower number of nodes --, and other processing steps that depend on location and context. Figure~\ref{fig:original-manual} provides an illustration thereof: the left panel (Figure~\ref{fig:original-manual}a) shows a street network prior to its simplification; the right panel (Figure~\ref{fig:original-manual}b) shows the same network after simplification, with several intersections, roundabouts, and dual carriageways removed, and the remaining geometries adjusted accordingly, ensuring continuity preservation. Note that the term ``simplification'' has been used in similar contexts with a slightly different meaning, such as smoothing the sinuosity of network edges (referred to as ``unfractalizing'' in \cite{jiang_fractal_2015}), the removal of interstitial nodes \citep{boeing_osmnx_2017}, or data storage format compression \citep{hendawi_road_2020}, among others. In the context of this paper, we explicitly define simplification as the process of converting the input network into its most simple morphological representation, as illustrated in Figure~\ref{fig:original-manual}.

\section{Previous work}\label{sec:previous-work}

The urgent need for an automated street network simplification method has been expressed within a striking variety of fields, such as urban morphology \citep{venerandi_form_2017, dibble_origin_2019}, bicycle network planning \citep{viero_bikedna_2023, ballo_modeling_2023, paulsen_societally_2023}, social data science \citep{aiello_urban_2025}, navigation in urban air space \citep{badea_limitations_2021, vidosavljevic_metropolis_2021}, and environmental modelling \citep{sanzana_decomposition_2018}. This corroborates the broad applicability and usefulness of the method proposed in this study. At the same time, given that this unresolved challenge appears across different disciplines, there is a lack of streamlined terminology for the problem at hand, which makes it particularly difficult to find relevant literature \citep{pueyo_overview_2019}. For example, \cite{lovelace_route_2024} refer to the same problem that we here call ``street network simplification'' as ``route network simplification.'' At the same time, similar terminology has been used for conceptually different problems: studies on map generalization refer to the problem as ``road network simplification'' \citep{hendawi_road_2020, baur_gradual_2022}, while the \texttt{OSMnx} package uses the term ``graph simplification'' when referring to the removal of interstitial nodes which are characteristic to raw OSM data \citep{boeing_osmnx_2017}. Below, we provide an overview of the most relevant previous studies working towards an automated street network simplification method, without claims to exhaustivity.

Most identified approaches are geometry-based. \cite{krenz_employing_2017} develops an algorithmic instruction of steps to follow, using several ArcGIS subroutines such as line simplification \citep{esri_simplify_2024}, in a simplification process explicitly tailored to OpenStreetMap data with corresponding attributes. Further, in their Python library \texttt{parenx}, \cite{deakin_anisotropi4parenx_2024} propose two morphological simplification approaches: skeletonization and Voronoi polygonization (see details below). 

Few graph-based simplification approaches could be found in previous literature. In an early study, \cite{jiang_structural_2004} propose a graph-based modelling approach to map generalization using a centrality-based node hierarchy filter. \cite{yang_detecting_2022} propose a graph-based machine learning approach for the detection of interchanges (i.e.~non-planar intersections) in street networks. Other graph-based \citep{jiang_scaling_2013} and entropy-based \citep{bjorke_map_2005} map generalization methods exist; however, the problem definition for map generalization is only partially overlapping with our working definition of simplification, and thus map generalization methods are not directly transferable to the street network simplification problem.

Several studies have combined methods from morphology and graph theory for their simplification workflows. As part of the \texttt{OSMnx} library, \cite{boeing_osmnx_2017} uses a user-defined distance threshold to merge all nodes within a distance from each other into single nodes, and reconnecting network edges while preserving the network topology. \cite{pung_road_2022} and \cite{cohen_complexity_2024} both propose iterative workflows that combine graph motives, graph metrics, geometry attributes, and the node consolidation method from \texttt{OSMnx}. In the Python package \texttt{cityseer}, \cite{simons_cityseer_2022} implements an iterative simplification workflow that includes both graph-based and morphological subroutines. Lastly, several toolboxes that enable (semi)automated simplification tailored to specific use cases or data contexts have recently become available, such as \texttt{SNman}, \texttt{Access Node Deriver}, and \texttt{OSMNetFusion} \citep{ballo_designing_2024, wohnsdorf_verkehrsplanung-und-verkehrsleittechnikaccess_node_deriver_2024, dahmen_osmnetfusion_2025}. 

\begin{table}[ht]
    \centering
    \begin{tabular}{cccccc}
         Method & OS & attr.-agnostic & case-agnostic & no manual & packaged \\
         \hline
         \cite{jiang_structural_2004} & \xmark & \cmark & \cmark & \cmark & \xmark \\
         \cite{krenz_employing_2017} & \xmark & \xmark & \cmark & \xmark & \xmark \\
         \cite{pung_road_2022} & \xmark & \xmark & \xmark & \cmark & \xmark \\
         \cite{cohen_complexity_2024} & \cmark & \xmark & \cmark & \cmark & \xmark \\
         \texttt{SNman} \citep{ballo_designing_2024} & \cmark & \xmark & \cmark & \xmark & \xmark \\
         \texttt{Access Node Deriver} \citep{wohnsdorf_verkehrsplanung-und-verkehrsleittechnikaccess_node_deriver_2024} & \xmark & \xmark & \xmark & \cmark & \xmark \\         
         \texttt{OSMNetFusion} \citep{dahmen_osmnetfusion_2025} & \cmark & \xmark & \cmark & \cmark & \xmark \\
         \hline
         \texttt{OSMnx} \citep{boeing_osmnx_2017} & \cmark & \cmark & \xmark & \cmark & \cmark \\
         \texttt{cityseer} \citep{simons_cityseer_2022} & \cmark & \cmark & \cmark & \cmark & \cmark \\
         \texttt{parenx} \citep{deakin_anisotropi4parenx_2024} & \cmark & \cmark & \cmark & \cmark & \cmark \\
         \hline
    \end{tabular}
    \caption{Overview of previous studies working towards an automated street network simplification method. 
    }
    \label{tab:methods-overview}
\end{table}

Ideally, the street network simplification method should fulfill five main requirements: it should be 1) open-source, 2) fully automated (i.e., not require any manual user interventions), 3) attribute-agnostic (i.e.~not requiring any attribute information on the geometries of the network), 4) case-agnostic (i.e., simultaneously addressing different subsets of the street network, such as roundabouts, complex intersections, and dual carriageways), and 5) fully packaged. In Table~\ref{tab:methods-overview}, we contrast the previous studies from the viewpoint of requirements met, and find that only three currently available methods -- \texttt{OSMnx} \citep{boeing_osmnx_2017}, \texttt{cityseer} \citep{simons_cityseer_2022}, and \texttt{parenx} \citep{deakin_anisotropi4parenx_2024} -- adequately meet this set of five core requirements. We review these three methods in more detail below, and subsequently use them to evaluate the output quality of our own method proposed here.

\subsubsection*{OSMnx}

The \texttt{OSMnx} package \citep{boeing_osmnx_2017} implements a method for intersection consolidation\footnote{\texttt{osmnx.simplification.consolidate\_intersections}}, described in detail in \cite{boeing2025topological}. This method merges all nodes within a certain buffer distance $d$ into one node and updates the affected network edges accordingly. The per-node buffer distance $d$ is customizeable for each individual node. However, \texttt{OSMnx} itself does not provide an estimation of the (locally varying) parameter $d$. This brings about the risk of introducing unwanted artifacts in places where the choice of $d$ is locally inappropriate. The intersection consolidation method works better for morphologically less complex intersections, and fails for morphologically more complex intersections, such as highway interchanges. In addition, since this method is targeted explicitly at intersection consolidation, it does not address other local cases of street network simplification\footnote{note that \texttt{OSMnx} uses the term ``to simplify'' differently, namely referring to the removal of interstitial nodes which are characteristic to raw OSM data, see the function \texttt{osmnx.simplification.simplify\_graph}}, such as dual carriageways.

\subsubsection*{cityseer}

The \texttt{cityseer} package \citep{simons_cityseer_2022} performs a complex, multi-step network simplification routine through a dedicated, internal helper function\footnote{\texttt{cityseer.tools.io.\_auto\_clean\_network()}} for the morphological simplification of street networks. This private function performs a sequence of 15 operations that generally follows an iterative ``remove-smooth-reconnect'' flow, which we describe in more detail in Appendix~\ref{appendix:cityseer}, and which notably includes steps for the merging of parallel edges (e.g.~divided highways). This routine, which is detailed in Section 5.1 of \citet{simons_cityseer_2022} and the \texttt{cityseer} Graph Cleaning Guide\footnote{\url{https://benchmark-urbanism.github.io/cityseer-examples/examples/graph\_cleaning.html\#manual-cleaning}}, is methodical and produces desirable results. However, users must pass in network data associated with tags from OSM for the function to work as intended. In our analysis -- see Section~\ref{sec:evaluation} -- we dissected the function to implement the process without relying on OSM tag attributes. This follows the spirit of \texttt{cityseer}-style simplification, while introducing as little bias as possible from our end.
 
\subsubsection*{parenx}

The \texttt{parenx} package \citep{deakin_anisotropi4parenx_2024} uses a very different approach compared to \texttt{cityseer}. Rather than implementing a routine dealing with individual issues, it implements two independent approaches, which both apply a single solution to the whole network based on its buffered version. The first approach is using raster-based image skeletonization to derive the simplified network from the overlapping buffers of the original geometry, while the second approach uses a Voronoi diagram to derive centerlines of the unioned buffer. Each of these requires further steps to link the original network data back to the simplified network. However, they are case-agnostic and can be used to process any linear network, e.g.~railways \citep{lovelace_route_2024}. 

\bigskip

\noindent In conclusion, the algorithmic simplification is clearly not a new problem researchers are trying to solve. However, none of the methods available in the literature and existing software are able to meet all of the five requirements stated above.  As a consequence, the applicability of available street network data is restricted to use cases where network representation is not causing issues, where partial simplification is acceptable, or where there are resources for costly manual simplification.

\section{Adaptive continuity-preserving simplification}\label{sec:new-way-to-simplify}

\begin{figure}[h]
    \centering
    \includegraphics[width=0.95\textwidth]{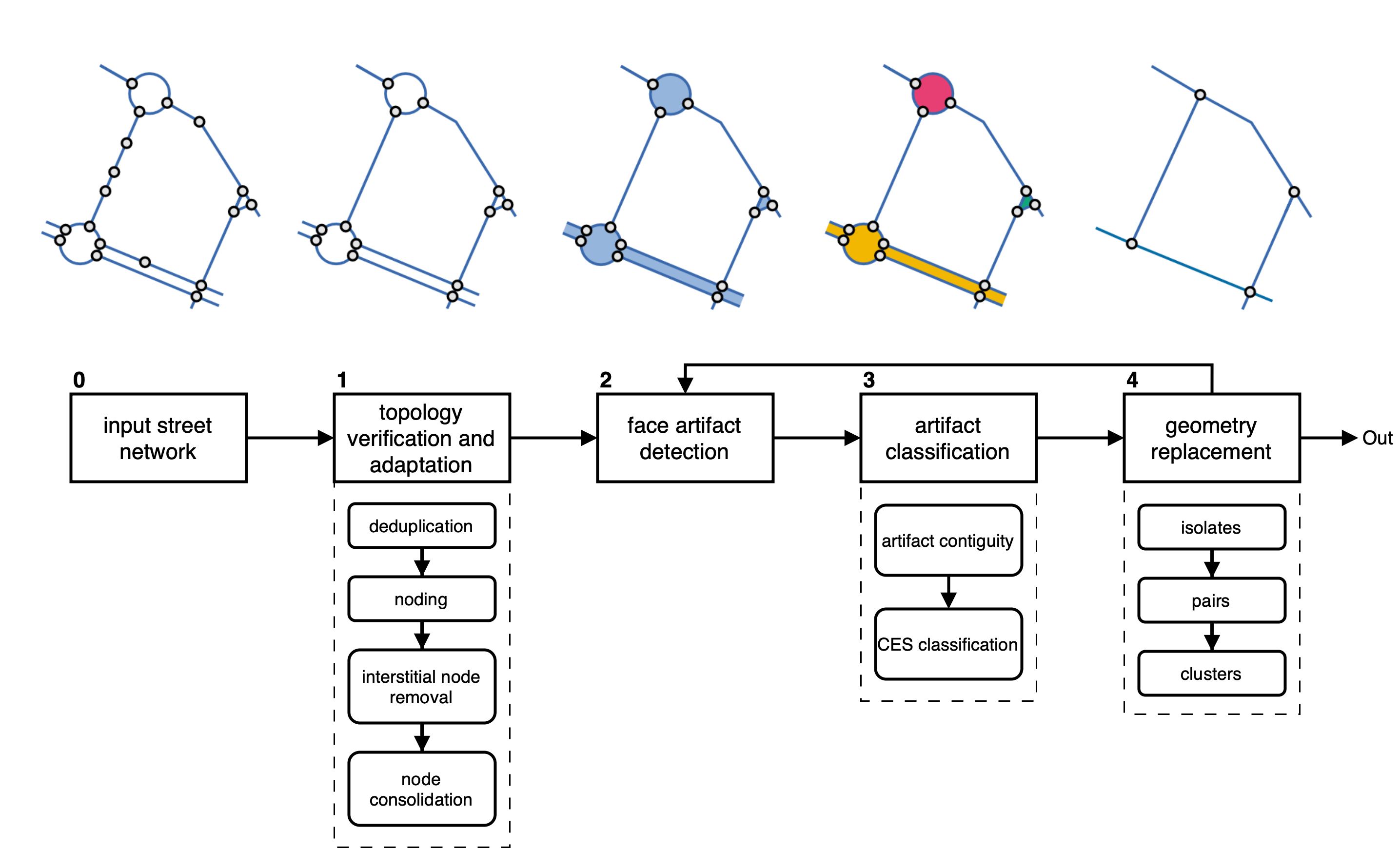}
    \caption{Conceptual diagram of the proposed simplification algorithm. Input street network (0) undergoes topological verification and adaptation (1) before face artifact detection (2). The face artifacts are then classified based on contiguity of polygons and continuity of edges forming them (3). The result goes through the geometry replacement pipeline that alters the existing geometry (4) based on the artifact classification to ensure minimal change and preservation of main network characteristics. Steps 2-4 are then repeated before returning the output consisting of a simplified street network.}
    \label{fig:algo-diagram}
\end{figure}

We propose a new, adaptive continuity-preserving simplification algorithm to resolve the simplification problem, while fulfilling all of the five requirements for the ideal method. The proposed algorithm consists of several steps, as shown in Figure~\ref{fig:algo-diagram}. The input data source is any street network represented as linear geometry (LineString geometry objects, if using the Simple Features specification \citep{sfa}). The attributes are not required or relevant to ensure generic applicability of the method to any input data source and to avoid dependency on potentially unreliable types of information (hierarchy tagging is inconsistent both across and within data sources \citep{witt_analysing_2021}).

The first simplification step is a verification of the topology of the network and a possible adaptation to fix the discovered problems. This step ensures that four requirements are met: 1) There are no missing nodes on geometries that touch via an endpoint of one of them -- the idea behind this is that if an edge ends at an intersection with another, there should be a node on both, as it is unlikely that such a situation represents a non-planar intersection; 2) there are no degree 2 nodes -- these have no meaning for the topology of the simplified network -- our framework assumes an edgeto be reflecting a street that either starts and ends at an intersection with another street or at a dead end; 3) there are no duplicated geometries, since duplicated LineStrings are usually an artifact of the data export and do not represent distinct edges; and 4) there are no nodes within a small tolerance (2 meters by default) of each other. 

The fourth requirement is solved by a consolidation of the network nodes. We use only a small tolerance (double the segmentation density used later in the pipeline) to eliminate nodes that are too close to each other to be considered individual nodes. The nodes that are within the tolerance of each other are merged and the street geometry is reconnected. The nodes to be merged are detected using the average linkage agglomerative algorithm to avoid over-consolidation. Average linkage ensures that there is no ``cluster chaining'' that would happen with algorithms like DBSCAN or similar and are present in node consolidation offered by \texttt{OSMnx}. The new node is defined as the weighted centroid of the cluster (computed as an average of the coordinates) to preserve the location of the majority of the existing connections. While we are using this algorithm to meet the topological criteria only, it is generalisable and can be used to consolidate nodes representing complex intersections into a single one if that is the only simplification step needed (e.g.~to measure node density).

The following steps form a loop that we iterate through twice. The loop starts with the detection of face artifacts using the method proposed by \cite{fleischmann_shape-based_2024}. This polygonizes the network and uses the relationship between the size and shape of the polygon to identify those that result from a traffic-oriented mapping, i.e., face artifacts. The original method is extended to allow more granular selection and additional detection of both false negative and false positive results. Polygons whose face artifact index is not less than the automatically derived threshold, but which touch the polygons identified as artifacts and whose shape characteristics are similar enough, are marked as artifacts during post-processing. In addition, the function can use an exclusion mask to mark all areas that should not be considered artifacts in any case (e.g.~an exclusion mask could be a layer of building polygons that captures the fact that a polygon containing a building is not an artifact, but a regular block). Face artifacts are then used to identify the portions of the network that require simplification. No other section of the network is altered by the algorithm. Given the artifacts are derived from the structure of each street network independently, the detection is naturally adaptive to different conditions ensuring that artifacts are detected correctly in various geographical and planning contexts with no need for any customised parameters.

The next step classifies all artifacts first into three groups based on artifact polygon contiguity and then into subgroups based on network continuity. The first level of classification divides all face artifacts into isolates (polygons with no neighbors), pairs (polygons with a single neighbor), and clusters (polygons with two or more neighbors). Isolates and pairs are further classified based on the number of edges forming the artifact, their continuity properties, and the location(s) of additional vertices.

The continuity-based classification is based on the algorithm proposed by \cite{tripathy_open-source_2021}, implemented in the Python package \texttt{momepy} \citep{fleischmann_momepy_2019}. The algorithm identifies continuity strokes as groups of edges that are geometrically continuous (the interior angle between them is above a certain threshold, typically 120 degrees). The original algorithm could potentially end a continuity stroke in the middle of an edge if there was a sharp angle. This violates the logic of network flow, where an activity can only end at a node. To overcome this limitation, we extended the original algorithm to include a so-called flow mode, which further restricts the detection of stroke groups by not allowing them to break in the middle of a geometry, forcing them to align with street network nodes.

\begin{figure}[h]
    \centering
    \includegraphics[width=0.95\textwidth]{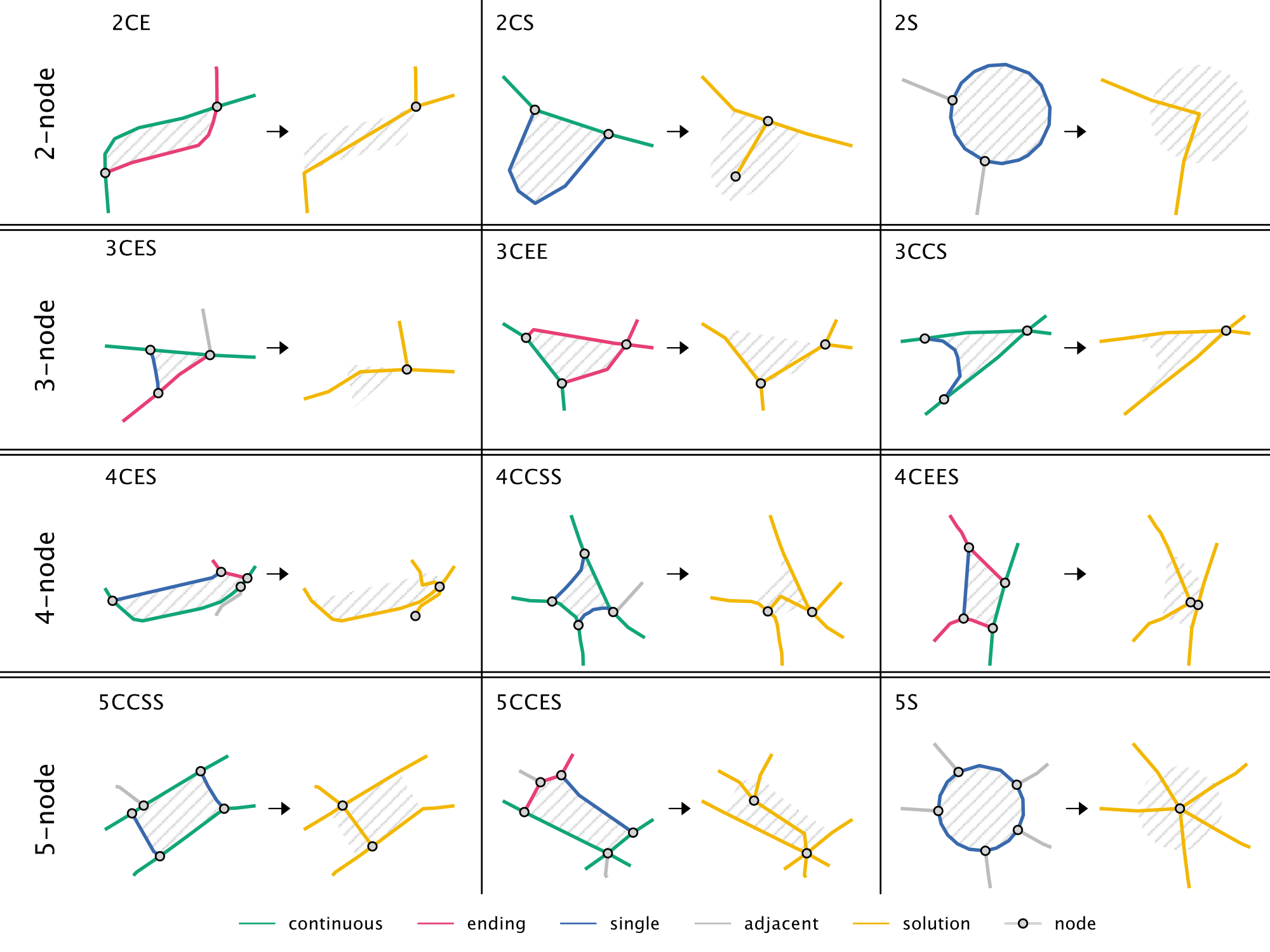}
    \caption{A subset of the CES typology derived from the Liège street network. A single CES type is composed of the number of nodes forming the artifact and the continuity types of strokes forming its boundary. For example, the most common type, 3CES, is composed of 3 nodes and 3 continuity strokes, one of each kind, while 5S is composed of 5 nodes and a single continuity stroke of type S (as it never leaves the artifact). The total number of CES types is theoretically infinite given no upper bound on number of nodes exists. The full overview of types present in Liège is available in Appendix \ref{appendix:ces}.}
    \label{fig:ces}
\end{figure}

Using the continuity stroke information, we can classify isolates and pairs into subgroups. The reason for this approach is to ensure that when a geometry is modified to remove the face artifact, the modification does not significantly alter the continuity of the resulting network. In other words, the simplification does not change its overall properties. From a continuity perspective, a single edge can be either \emph{continuous} (C), meaning that its stroke continues both before and after the edge, \emph{ending} (E), when its stroke continues only on one end but not the other, or \emph{single} (S), composed of edges that are their own stroke group. The resulting CES classification, illustrated in Figure \ref{fig:ces}, ensures that the simplification primarily affects the geometry of S, followed by E, and only then C, ensuring that in most cases the C edges remain unchanged and the entire stroke remains largely intact. To further guide the simplification, the CES classification captures locations of additional vertices, to distinguish between those subdividing one C stroke into two edges and those subdividing E or S. A more detailed overview of the classification and proposed solutions for each type is available in Appendix~\ref{appendix:ces}.

The next step in the algorithm is geometry replacement. For isolates, this follows a heuristic derived from the CES classification described in the previous paragraph. For pairs, it either identifies the edge connecting the two polygons as \emph{continuous}, in which case it treats each part as an isolate to preserve continuity, or it removes the common edge and applies the CES classification to the entire artifact pair, using the aforementioned heuristics. Clusters are first merged into a single polygon per cluster and then replaced by a polygon skeleton derived from a Voronoi diagram of their edges. Each of the modified geometries is then marked as either \emph{new}, if it is completely generated during the process, or \emph{extended}, if it is mostly the original geometry that has been modified to ensure reconnection of the mesh.

In the last algorithm step, the resulting simplified network is then used as input for the second loop of the algorithm, which starts with the new artifact detection using the artifact threshold already determined from the first loop and ends with the topological post-processing. The second loop is necessary because occasionally the first loop may produce new artifacts that need to be resolved. However, this is typically a matter of only a handful of cases.

\section{Evaluation}\label{sec:evaluation}

The evaluation of the proposed method is done through the comparison of its outcome to those of the state-of-the-art simplification algorithms that fit the following three criteria: 1) are attribute-agnostic, meaning that no expected structure of attributes is needed. This criterion eliminates methods tied to, for example, OpenStreetMap data like \texttt{SNMan}; 2) attempt to simplify both nodes and edges, meaning that the algorithm needs to attempt to resolve both ``clusters'' of nodes representing complex intersections and multi-line carriageways. This criterion eliminates methods focused solely on node consolidation like those available in \texttt{momepy}; and 3) are based on open software, meaning that no purchased license or formal approval is needed to run the software. This eliminates tools like those offered in the products by Esri (e.g.~ArcPro, although as noted above those do not reach the state-of-the-art quality anyway). These criteria resulted in the following set of tools used for comparison: \texttt{cityseer}, \texttt{OSMnx}, and \texttt{parenx} with both methods offered -- skeletonize and Voronoi.

While each of the tools offer its own set of parameters that allow fine-tuning the implementation to a specific network, we use the default parameters only. Case-specific parameters are not possible when the analysis deals with a large number of cities from various regions. In such a case, it is necessary for the algorithm to either work with the default parameters or to automatically adjust as is the case with \texttt{neatnet}'s implementation, in which the underlying face artifact detection is always case-specific and provides intrinsically adaptative parameters of the algorithm.

The core evaluation of the performance is based on the comparison of the algorithmically simplified networks to the version of the simplified networks resulting from a manual, hand-edited process. The performance is measured in a set of seven street networks sampled from cities across the globe, to reflect the geographical heterogeneity of urban development. This selection is a subset of the cities used in \cite{fleischmann_shape-based_2024} for consistency with their original publication. The evaluation is done both quantitatively, by measuring the similarity of the result to the manual ``ground truth'' data reflecting the algorithms' simplification performance, and qualitatively, by visual inspection and comparison of the results. In addition, we report computational efficiency (runtime speed and peak memory consumption) allowing assessment of application feasibility for each of the methods. Computational efficiency is measured within a single frozen reproducible Python environment on a single machine running Ubuntu 22.04.

The quantitative similarity-based performance is measured based on the overlaid H3 grid of resolution 9, with a hexagonal edge length approximately 200 meters\footnote{\url{https://h3geo.org/docs/core-library/restable\#edge-lengths}}. The evaluation occurs on the basis of individual grid cells where each metric is measured independently of the other cells. This allows granular understanding of potential spatial heterogeneity of the algorithms' performance. We capture a set of metrics reflecting the structure of the network: 1) average node degree within the cell; 2) total number of coordinates required to encode the geometries in the cell; 3) number of edges intersecting the cell; 4) total length of the network within the cell; 5) number of continuity strokes intersecting the cell; 6) maximum length of continuity strokes intersecting the cell; and 7) total length of continuity strokes intersecting the cell. 

Based on the results, we measure the difference in metrics between the original network and the manually simplified one using two methods: Euclidean distance and correlation. Let $p, q$ be respective distributions of difference in two methods being compared, where $p_{i}$ and $q_{i}$ are the values of the respective distributions for grid cell $i$. Then the difference $d$ is measured as the Euclidean distance between the two distributions defined in Eq.~\ref{eq:distribution_difference_euclidean} as:

\begin{equation}\label{eq:distribution_difference_euclidean}
d(p, q) = \sqrt{\sum_{i=1}^{n} (q_i - p_i)^2}
\end{equation}

\noindent Furthermore, we capture the correlation using the Chatterjee's $\xi$ coefficient \citep{Chatterjee02102021}, which proves more effective to distinguish between the methods than standard Pearson or Spearman correlations in this specific case (see Appendix \ref{appendix:corr}). For the set of grid cell value pairs $(p_{i},q_{i})$, ordered by ascending value of $p_{i}$, let $r_i$ be the rank of $q_i$; then the coefficient is defined in Eq.~\ref{eq:chatterjee_xi} as:

\begin{equation}\label{eq:chatterjee_xi}
\xi_n(p, q) := 1 - \frac{3 \sum_{i=1}^{n-1} |r_{i+1} - r_i|}{n^2 - 1}
\end{equation}

Visual inspection (Section \ref{sec:vis-inspect}) is based on the comparison of the original, manual, and algorithmically simplified networks with a specific focus on areas of significant differences reported by the grid-based metrics.

\section{Results}\label{sec:results}

Below, we report evaluation results for computational efficiency and simplification performance, as well as comparison which allows a direct evaluation of the outcome. 

\subsection{Computational efficiency}

Computational efficiency of individual simplification algorithms within each of the case studies is reported in Table~\ref{tab:runtime_memory}. The differences in runtime are significant with the slowest approach (\texttt{parenx} Voronoi) taking over 165x longer than the fastest one (\texttt{OSMnx}). However, it also needs to be considered that not every method attempts to do the same thing, with \texttt{OSMnx} effectively not simplifying dual carriageways and other similar components of street networks. \texttt{neatnet} sits in the middle with a median runtime of $57.5s$. An analogous situation is observed in memory consumption, where \texttt{parenx} (skeletonize) tends to require particularly high amounts of RAM.

The longer runtimes and dedicated memory burden can be mostly explained by the usage of Voronoi diagrams (and similar). Since \texttt{neatnet} relies on generating Voronoi diagrams for small subsets of study areas, its runtime and memory burden is generally larger than that of \texttt{cityseer} and \texttt{OSMnx} (which do not make use of Voronoi diagrams), but dramatically smaller than the two \texttt{parenx} methods (which apply them across the whole study area).

Even when computational efficency is not considered as crucial as other evaluation criteria, it is clear that some methods require substantially more resources to run, which may not be generally available.

\begin{table}[h!]
\centering
\sisetup{mode = text, reset-text-shape = false}
\begin{tabular}{lrSSSSS}
& & \texttt{cityseer} & \texttt{OSMnx} & {\texttt{parenx} (Voronoi)} & {\texttt{parenx} (skelet.)} & \texttt{neatnet} \\
\midrule
\multirow{7}{*}{Runtime [s]} & Aleppo & 26.3 & 10.2 & 1922.3 & 398.1 & 137.1 \\
 & Auckland & 13.8 & 6.2 & 436.6 & 100.5 & 41.4  \\
 & Bucaramanga & 17.3 & 9.0 & 3070.9  & 127.1 & 55.5 \\
 & Douala & 23.8 & 10.9 & 1480.8 & 162.0  & 76.2 \\
 & Liège & 18.4 & 8.4 & 1014.3 & 324.6  & 57.5 \\
 & Salt Lake City & 12.4 & 6.0 & 1023.9 & 163.5 & 45.0 \\
 & Wuhan & 26.7 & 9.6 & 3972.9 & 2323.5  & 368.5 \\
 \cline{3-7}
& \textit{median} & \itshape 18.4 & \itshape 9.0 & \itshape 1480.8 & \itshape 163.5 & \itshape 57.5 \\
\midrule
\multirow{7}{*}{Memory [MiB]} & Aleppo & 75.09  & 145.62 & 14855.31 & 45252.99 & 3157.28 \\
 & Auckland & 25.25 & 32.03  & 4214.23 &  14905.45 & 3.80 \\
 & Bucaramanga & 91.66 & 103.28  & 16964.75 & 84164.83  & 151.97 \\
 & Douala & 109.69 & 149.38 & 9088.87 & 15788.22 & 1867.03 \\
 & Liège & 97.50   & 91.09 & 7474.78 & 34269.82 & 141.54  \\
 & Salt Lake City & 21.48 & 18.91 & 8728.17 & 18723.09 & 246.971 \\
 & Wuhan & 120.94  & 62.97 & 27913.33 & 118855.29 & 4793.41 \\
 \cline{3-7}
& \textit{median} & \itshape 91.66 & \itshape 91.09 & \itshape 9088.87 & \itshape 34269.82 & \itshape 246.97 \\
\bottomrule
\end{tabular}
\caption{Computational efficiency of different methods.}
\label{tab:runtime_memory}
\end{table}

\subsection{Simplification performance}\label{sec:simp-perf}

\begin{figure}[h]
    \centering
    \includegraphics[width=0.95\textwidth]{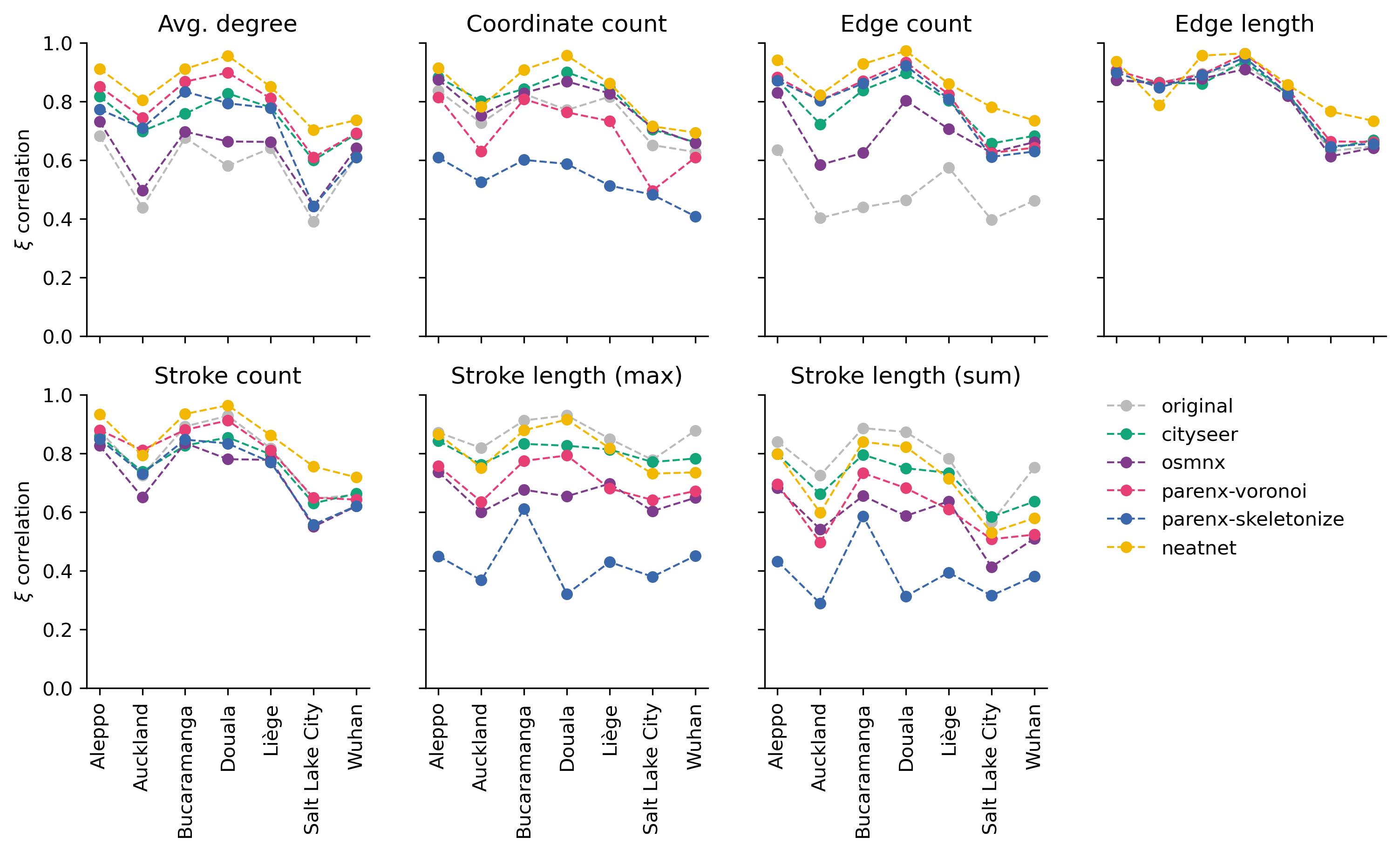}
    \caption{Chatterjee's $\xi$ correlation coefficient between properties of manually simplified networks and networks based on each of the tested algorithms vs.~the original network as a baseline. Higher is considered better -- approaching $1.0$ on the $y$-axis.}
    \label{fig:chi}
\end{figure}

To evaluate simplification performance, we compute the deviation of the street network simplified by each of the tested methods from the one simplified manually, which is considered the ``ground truth'' in this scenario, for seven different metrics (average node degree, coordinate count, edge count, total edge length, and three continuity-based metrics described in detail below), in a gridcell-by-gridcell approach (H3 of resolution 9). Figure~\ref{fig:chi} shows the outcome of the $\xi$ correlation coefficient, which would approach a value of $1.0$ for a network perfectly matching the manual one. In this sense, the higher the value of $\xi$, the better the performance of the corresponding method. For reference, we also report the correlation between the original input data and the ground truth.

Starting with the average node degree, it is clear that \texttt{OSMnx} underperforms across all study areas when compared to other methods, closely mirroring the average node degree of the original, unsimplified, network data. That indicates that the algorithm is rather conservative when it comes to changes and leaves many artifacts intact, possibly due to the default parameters being insufficient for the selected use cases. On the other hand, \texttt{neatnet} outperforms the other methods in all study areas, making it the most similar to the ground truth.

Coordinate count, which captures how many points are required to represent the network, indicates that both \texttt{parenx} methods appear to generate results that are actually less correlated than the original data, making the outcome worse than the input in this sense (introducing additional points in the data and thus increasing coordinate count). Again, \texttt{neatnet} performs at least as well as the other methods in all scenarios except \texttt{cityseer} in Auckland.

Edge count shows the highest deviation between original and manual data, leaving the most space for change. Similarly to average degree, which is conceptually related, \texttt{OSMnx} underperforms across all study areas except Salt Lake City and Wuhan where it performs roughly equivalently to the \texttt{parenx} methods and \texttt{cityseer}. \texttt{neatnet} outperforms all other methods in all study areas.

Total edge length, on the other hand, shows the least space for change. All the methods perform roughly equivalently, with \texttt{neatnet} slightly outperforming other methods in all scenarios except Auckland, where it is the worst.

The three remaining metrics -- stroke count, length of longest stroke, and sum of stroke lengths -- are derived from the assessment of continuity of networks, with the underlying rationale of ensuring that the simplification preserves continuity properties as closely as possible. Stroke count, which captures the number of continuity strokes per H3 cell, shows that all methods perform roughly equivalently with \texttt{neatnet} slightly outperforming other methods in all scenarios except Auckland, where \texttt{parenx} skeletonize is better. Both maximum length and total sum of lengths of strokes crossing each H3 cell show similar behaviour, where the manually simplified network is closest to the original one. In this context, \texttt{neatnet} and \texttt{cityseer} perform similarly and both are generally slightly less correlated to the manually simplified network than the original network data, with the other methods performing worse.

\begin{figure}[h]
    \centering
    \includegraphics[width=0.95\textwidth]{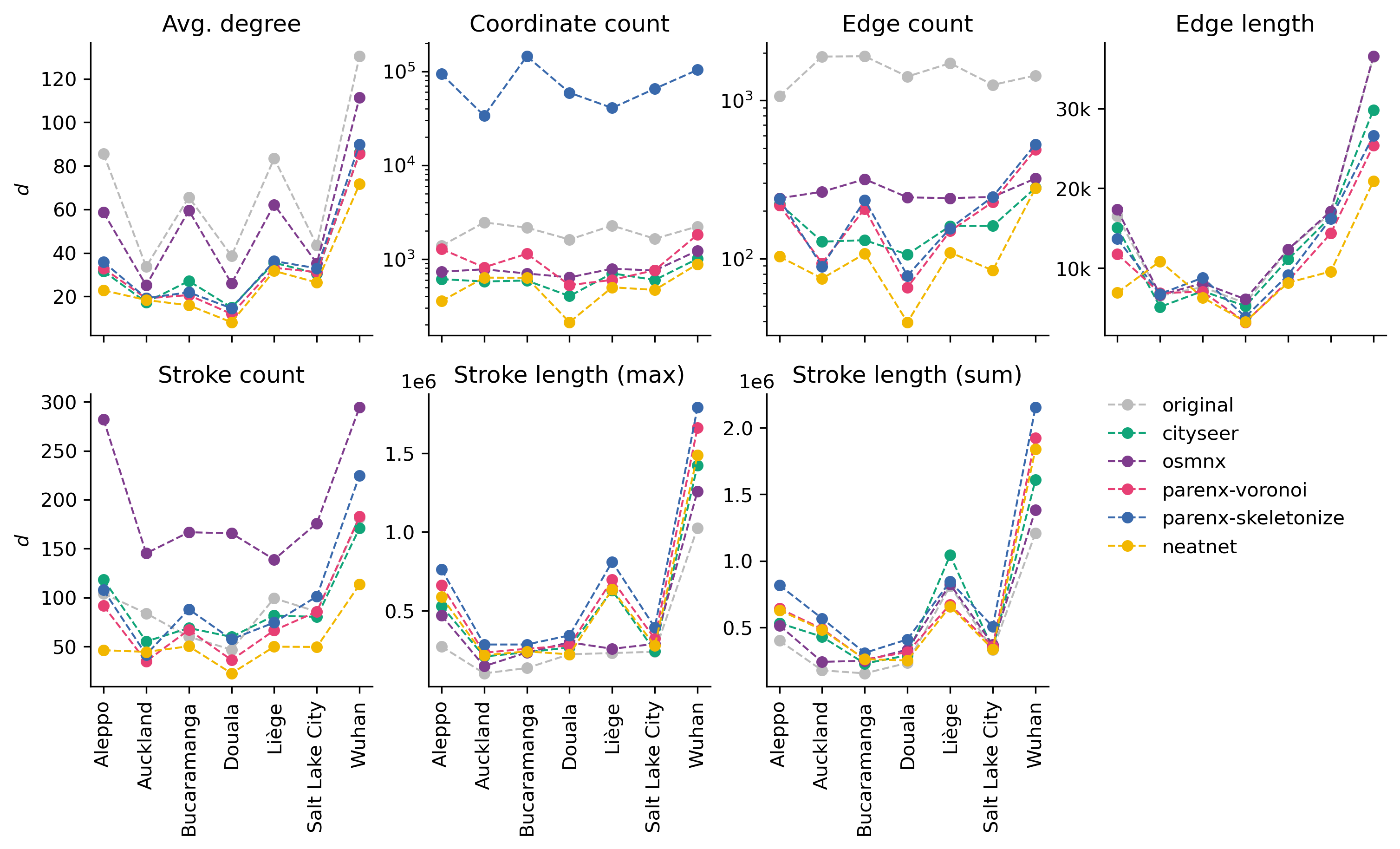}
    \caption{Euclidean distance between properties of manually simplified networks and networks based on each of the tested algorithms vs.~the original network as a baseline. Lower is considered better -- approaching $0.0$ on the $y$-axis.}
    \label{fig:delta}
\end{figure}

The same data may also be evaluated not only via correlation but using the absolute deviation capturing the Euclidean distance between two distributions, shown in the Figure~\ref{fig:delta}. We are aiming to minimise the deviation, i.e., the lower its value, the better the simplification perfomance.

At a high level, we can see that the relative performance practically mirrors the one observed in the correlation analysis above, with \texttt{neatnet} generally outperforming the other methods. However, the deviation provides another dimension to the interpretation, as it indicates the magnitude of the error. This is worth pointing out especially in the case of coordinate and edge counts, and the values reported by \texttt{parenx-skeletonize} and the original data, respectively. The skeletonization algorithm is based on rasterisation and without further simplification, which is not enabled by default, \texttt{parenx} tends to produce orders of magnitude more coordinates than the other methods, making the geometries very dense. The high deviation of the edge count, on the other hand, captures the initial situation where the network contains a high number of interstitial nodes, splitting single edges that would typically connect two intersections into multiple ones.

Combining the observations from the simplification performance assessments, \texttt{neatnet} proves to be the consistent top-performer over the other methods with exceptions of individual case and metric combinations, where other methods may occasionally be better.

\subsection{Visual inspection}\label{sec:vis-inspect}

\begin{figure}[h]
    \centering
    \includegraphics[width=0.95\textwidth]{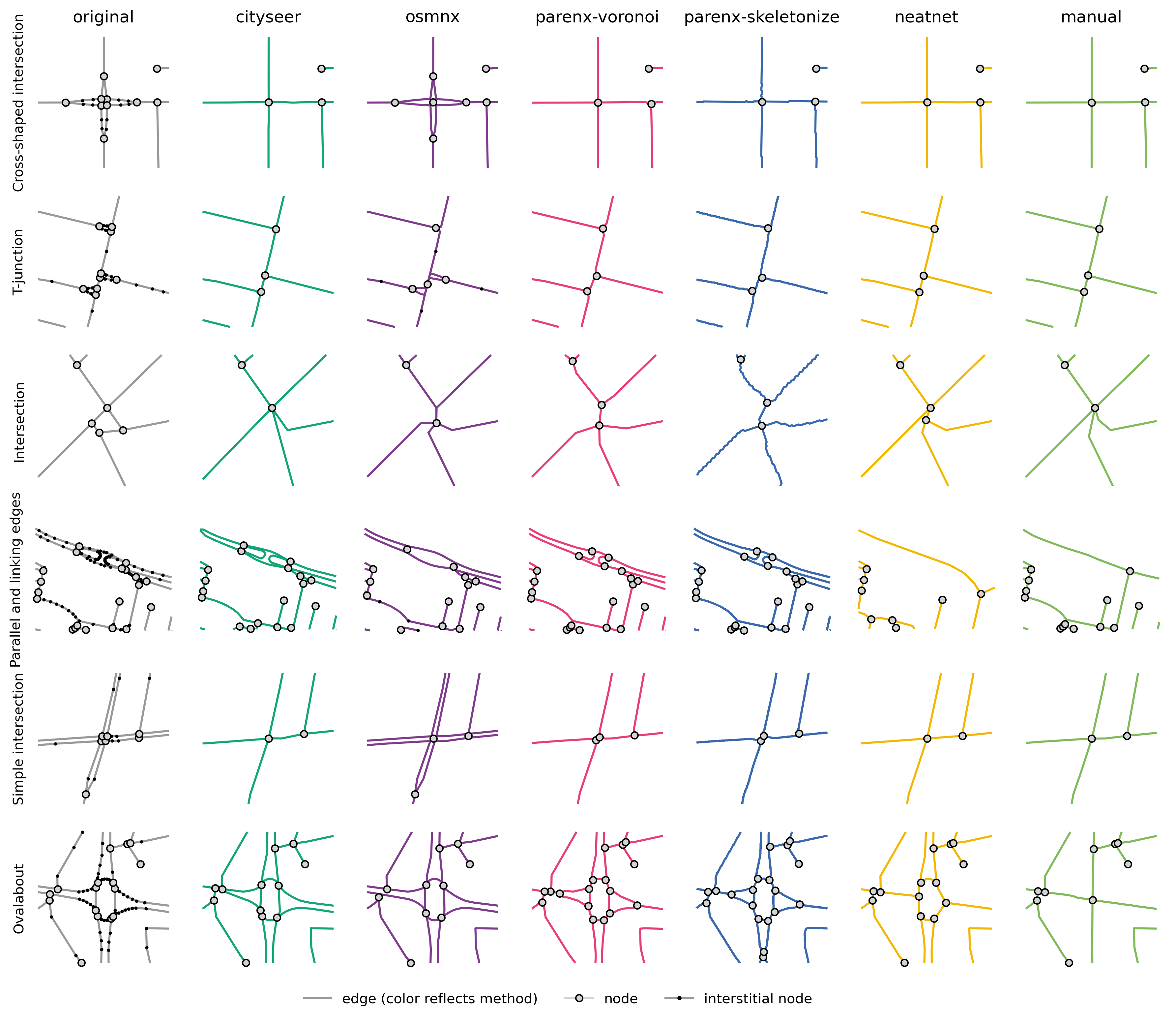}
    \caption{A small set of example situations that ought to be simplified and the resutling simplified networks from all tested methods, together with the comparison to the manual simplification considered ground truth.}
    \label{fig:situations-selection}
\end{figure}

Figure \ref{fig:situations-selection} contains a small subset of situations that require some degree of simplification and a visualisation of how each of the methods deal with them (see Appendix \ref{appendix:protocol} for a manual simplification protocal with a larger set of example situations; and Appendix \ref{appendix:solutions} for an illustration of how each of the methods deals with each of the example situations). This form of visual inspection helps us to provide necessary context to the numeric evaluation in the previous section and another level of understanding of how each of the methods perform. Note that the actual visual inspection task has been done based on whole networks.

We can draw some general conclusions, but there are many differences at individual intersections and other parts of the networks. Overall, the story remains largely the same as the numerical evaluation already suggested. Both \texttt{neatnet} and \texttt{cityseer} tend to do a relatively good job in most places. The main difference lies in large highway intersections, where \texttt{cityseer} typically retains the transporation geometry as is, while \texttt{neatnet} attempts to simplify (see the cases ``Parallel and linking edges,'' and ``Ovalabout''), albeit not always as cleanly as the manual simplification does. \texttt{parenx} methods tend to result in similar networks (rather unsurprisingly due to underlying algorithms), with the main difference being the density of vertices capturing the processing grid leading to orthogonal edges in the skeletonize approach, mentioned in the previous section. Besides that, the networks are relatively well simplified, though with occasional remapping of street continuity (see case ``Intersection''). \texttt{OSMnx} shows the most limited outcomes, with node consolidation occasionally producing partially overlapping geometries leading to topological simplification but not geometric one (e.g.~case ``T-junction''). Additional illustrations are available in Appendices \ref{appendix:vis-inspection} and \ref{appendix:solutions}. For better illustration, it is recommended to look through the complete data in an interactive session. See Section \ref{sec:datacode} for data access.

\section{Discussion}\label{sec:discussion}

Our novel adaptive continuity-preserving simplification algorithm, released within the open-source \texttt{neatnet} package, proves to be a robust option for street network simplification. While it is not the fastest or the most memory-efficient among the tested options, it compensates computational demands by consistently providing the best simplification outcome across our case studies. The key advantage over the other implementations is the focus on continuity preservation, ensuring that the simplification produces naturally flowing geometries, avoiding undesired relocations of intersections and their adjacent edges. Allowing for one or more of the simplification requirements outlined in Section~\ref{sec:previous-work} (open-source, fully automated, attribute-agnostic, case-agnostic, and fully packaged) to be violated, the results could have been different. However, when following all five requirements, our results suggest that \texttt{neatnet} is the best option currently available.

Unless the use case focuses solely on geometric or network properties of resulting simplified networks, there is a clear value in attribute preservation between the original and simplified data. In this regard, \texttt{cityseer} and \texttt{OSMnx} do not alter the attributes of geometries that are not causing artifacts and are able to partially aggregate data in some steps (e.g.~removing interstitial nodes). However, when dropping dual edges, attributes are retained for the shortest geometry only. \texttt{neatnet} does not alter the attributes of geometries external to the artifacts either and can equally aggregate data when removing interstitial nodes. However, the newly generated connections (typically when resolving face artifact clusters) are left without attribute data as the mapping between the original and generated geometries is undefined. \texttt{parenx}, due to their approach of treating the whole network as potentially worth simplification, does not preserve any attributes. However, the same authors have developed a library -- \texttt{ANIME} \citep{parry_josiahparryanime_2025} -- that can perform the data transfer based on a partial matching algorithm ex-post (applies to all the outputs, not just \texttt{parenx}), making the in-simplification preservation less critical.

Apart from the simplification performance, there are some additional advantages of the \texttt{neatnet} package and the implementation details of the algorithm. The first is its simplicity: to run the adaptive continuity-preserving simplification algorithm, a user needs a single function, typically with one single argument specifying the input street network. There is no need to consider the sequence of individual simplification steps required in \texttt{OSMnx} and even more in \texttt{cityseer}. However, the algorithm remains configurable, even though it is not typically needed as the underlying face artifact detection intrinsically adapts to the configuration of the local network. When compared to the majority of scenarios, special use cases (e.g.~railway networks) or network configurations may have slightly varying needs, and some threshold adaptation may be needed for artifact detection. All parameters that influence what is being simplified are exposed and can be altered if needed. This also includes a special argument enabling a user to pass an explicit collection of geometries indicating which locations should not, by any means, be considered artifacts. As discussed in detail in \cite{fleischmann_shape-based_2024}, the face artifact detection routine produces a small subset of false positives. In urban environments, these false positives typically contain buildings or water bodies (e.g.~Amsterdam's canal system) and a related geometry layer can be passed to \texttt{neatnet} as a so-called exclusion mask, ensuring that no polygon that contains a geometry will be treated as an artifact. The false positive rate is relatively small and its effect on the simplification performance reported in this paper is analysed in the Appendix~\ref{appendix:eval-with-buildings} by using buildings as an exclusion mask.

Despite the new algorithm outperforming the existing state-of-the-art solutions, it is not perfect. Besides the occasional false positives in artifact detection, the large non-planar highways and interchanges may occasionally be simplified in a way that alters the topology of the original network (typically by allowing more directional changes than the original). However, for the typical use of a morphological network, this is not necessarily of interest as what matters is the development surrounded by buildings. A minor known issue also lies in the simplification of extremely concave artifacts. When the CES typology suggests replacement with a centreline, this may be generated outside of the original artifact. In such a case, the algorithm does not return the new connection due to the strict requirement of these being contained by the artifact and warns that the artifact was insufficiently simplified. Nevertheless, these cases are rare. 

Beyond the known algorithmic limitations, the analysis presented in this paper is also limited to the scope of the work. For one, we are testing only a single source of data, OpenStreetMap networks. While we do not expect it, street network data coming from other sources (e.g.~national mapping agencies) might showcase different simplification needs affecting the performance of each of the method. Also, it should be noted that the results shown in the previous section are not necessarily the best outcomes the methods are capable of. When we take an iterative manual effort to fine-tune the default parameters, both \texttt{OSMnx} and \texttt{cityseer} will likely be able to reach lower deviations from the manual ground truth data. Yet, that would violate one of the requirements for the algorithm -- full automation. Similarly, both can make use of the OSM tags, when those are available and reliable. However, this feature is significantly limited in its  applicability, as the data may come from various sources and even the information on hierarchy may not be directly mappable to the one used by OSM, breaching another of the requirements -- being attribute-agnostic.

We have put forth a thorough framework for dealing with all major and most edge cases observed in modern urban areas. While we can claim our CES classification schema and methodology are rigorous, we cannot claim that it is absolutely exhaustive. Therefore, one avenue of future work includes regular review and addition of results in new study areas to continually hone the algorithm's performance in dealing with atypical scenarios. More work remains to be done on improving the computational performance of the procedure by moving some of the code from its current Python implementation to more performant compiled code (e.g.~using just-in-time compilation or a properly compiled language like Rust) and by improving the internal logic of the existing code. Nonetheless, the main focus of future work shall focus on the application of the algorithm not for the sake of simplification but as a pre-processing step in a larger processing pipeline, focusing on street network analysis, morphological assessments, or data visualisation, to validate its applicability in real-world scenarios.

In this work, we are using one perspective on what simplification is and how it should be conceptualised, focusing on preserving configurative and continuity properties of the networks. We acknowledge that some may view the definition differently and desire to expand it. That may be the case of simplification for map generalisation purposes \citep{mackaness_map_2014}, where the aim is not only to collapse geometries representing the same street or intersection into a single one, but also to remove the geometries of lower hierarchies. We argue that while our definition and implementation does not aim to fulfill the needs of map generalisation, it could form one step in the generalisation pipeline which first simplifies the network and then drops the geometries of lower hierarchies.

Street networks simplified using the adaptive continuity-preserving algorithm avoid the consequences of using transportation-focused networks for non-transportation purposes: It could actually completely resolve the issue seen in \cite{araldi2024multi}. Their morphological assessment is sensitive to directional changes in the geometry, hence shares the requirements on continuity preservation posed to the work presented here. Similarly, the work on urban blocks \citep{louf_typology_2014} would avoid complicated post-processing of generated block polygons as the entirely simplified network should not, in theory, produce any that do not represent urban blocks. The same applies to routing and navigation studies, where a simplified network minimises the misallocation of GPS tracks to streets.

The breadth of disciplines that could benefit from the work presented in this paper goes beyond the original motivation to generate networks that could support morphological analysis. In analyses that require generic routing (not linked to one-way streets and such), simplified networks could significantly reduce the computational demands and works like \cite{knaap_segregated_2024} linking network connectivity to segregation could rely on network topology metrics not affected by transportation geometry. Similarly, Space Syntax approaches which originally depended on axial maps \citep{hillier_space_1996} but is increasingly using street network data \citep{serra_angular_2019}, shall benefit here as the simplified network is conceptually as close to the topological representation of urban space as possible. Data visualisation needs expressed by \cite{lovelace_route_2024} or drone routing requirements by \cite{morfin2023u} are two more applications that may benefit, among many other use cases. 

In conclusion, in this manuscript we introduced a novel method for street network simplification. Our approach is automated and adaptive, and integrates both morphological and graph-based approaches to achieve results that mimic manual, point-and-click simplification to a high degree. Moreover, network edge attributes (if present) are preserved and flow continuity is retained to perpetuate the original physical character of the street network. While our algorithm \textit{does not} outperform the other state-of-the-art approaches universally (i.e., across all study areas and all comparative measures), it \textit{does} outperform the other state-of-the-art approaches in the vast majority of scenarios. While the algorithm presented here is both runtime and computationally efficient, future work, enabled by our open-source framework, could further improve the logic and performance of our method. 

\section*{Data and code availability}\label{sec:datacode}

The whole method is encapsulated in a series of Jupyter notebooks executed in a locked Pixi environment, ensuring full reproducibility. All components of the work rely on open source software and open data, with the resulting code and data being openly available at \url{https://github.com/uscuni/simplification}
and archived at \url{doi.org/10.5281/zenodo.15263789}. The simplification routines have been released as the open source package \texttt{neatnet}, available at \url{https://github.com/uscuni/neatnet} and archived at \url{https://doi.org/10.5281/zenodo.14765801}.

\section*{Acknowledgements}

The authors kindly acknowledge funding by the Charles University’s Primus programme through the project ``Influence of Socioeconomic and Cultural Factors on Urban Structure in Central Europe'', project reference PRIMUS/24/SCI/023. \\

\noindent This research could not have been done without the open-source community and OpenStreetMap contributors. We would like to thank Tomáš Hanula and Jiří Jakubíček for their help in manual simplification. Further, we would like to thank \href{https://www.ornl.gov/staff-profile/daniel-s-adams}{Dr. Daniel S. Adams} for his prelimiary review of this manuscript and value feedback and to all the people who have taken part in discussions about the idea and its implementation, among which we would like to name Krasen Samardzhiev and Lisa Winkler. \\

\noindent Copyright: This manuscript has been authored in part by UT-Battelle, LLC, under contract DE-AC05-00OR22725 with the US Department of Energy (DOE). The publisher acknowledges the US government license to provide public access under the DOE Public Access Plan (\url{http://energy.gov/downloads/doe-public-access-plan}). \\

\bibliography{references}

\clearpage

\appendix

\section*{Appendices}

\section{Manual simplification protocol with use cases}\label{appendix:protocol}

Table~\ref{tab:situations_and_descriptions} details 19 cases as a protocol for manual simplification. Included are, by column,~(1)~a situation label,~(2)~a situation description,~(3)~a location from our evaluation data set where an example of the situation can be found,~(4)~a plot of the situation (before simplification), and~(5)~a plot of the solution (after simplification). While we cannot claim general exhaustivity of this list, the situations listed here cover all cases that have occurred in our evaluation data set. Appendix~\ref{appendix:solutions} contains illustrations of simplified solutions, by method, for each use case from Table~\ref{tab:situations_and_descriptions}.

\begin{longtable}{p{1.8cm}p{3cm}p{1.3cm}p{2.5cm}p{2.5cm}}
\caption{Use cases for manual simplification protocol.} \\
\hline
\textbf{Situation} & \textbf{Description} & \textbf{Example} & \textbf{Before} & \textbf{Goal} \\
\hline
\endfirsthead

\hline
\textbf{Situation} & \textbf{Description} & \textbf{Examples} & \textbf{Before} & \textbf{Goal} \\
\hline
\endhead

\hline
\endfoot

\hline
\endlastfoot

Parallel edges & Collapse the parallel edges, but be careful not to create a connection that does not exist (e.g.~ there is a wide barrier between the edges, which cannot be crossed). & 9.81981°E 4.00249°N & \adjustbox{valign=t}{\includegraphics[width=2.5cm]{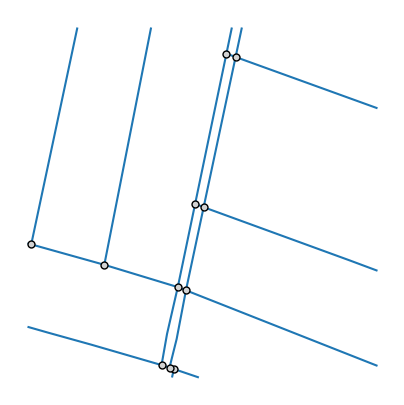}} & \adjustbox{valign=t}{\includegraphics[width=2.5cm]{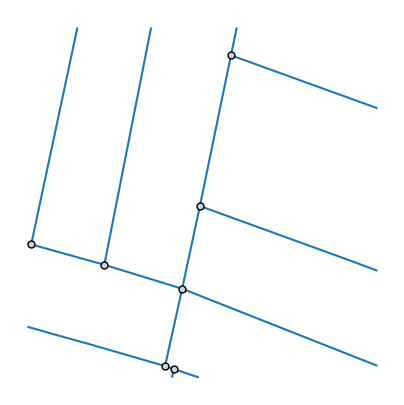}}\\
Roundabouts & Delete the roundabout and create an intersection. The edges will intersect in the center of the original roundabout. & 9.82132°E 4.00986°N & \adjustbox{valign=t}{\includegraphics[width=2.5cm]{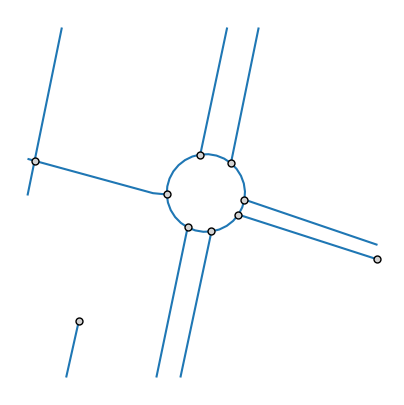}} & \adjustbox{valign=t}{\includegraphics[width=2.5cm]{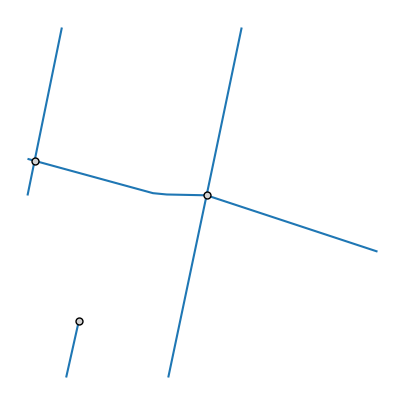}} \\
Diverging streets & Streets meeting at an intersection, can be interconnected with side streets. Identify the main street and delete the side streets. & 37.06837°E 36.31436°N & \adjustbox{valign=t}{\includegraphics[width=2.5cm]{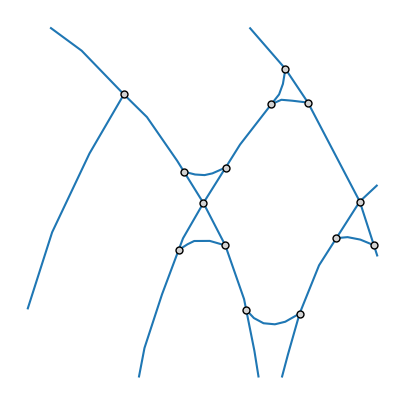}} & \adjustbox{valign=t}{\includegraphics[width=2.5cm]{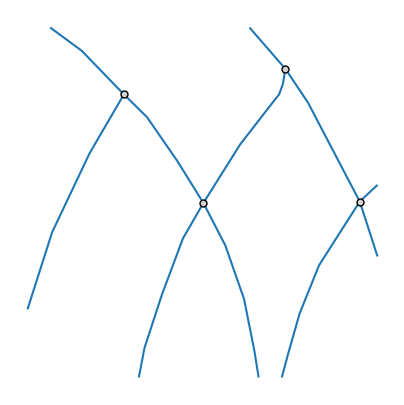}} \\
T-junction & Street diverging before an intersection. Delete the diverging streets and snap the point of divergence on the intersecting street. & 174.75120°E 36.88972°S & \adjustbox{valign=t}{\includegraphics[width=2.5cm]{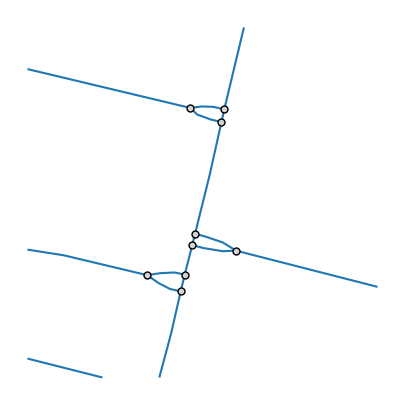}} & \adjustbox{valign=t}{\includegraphics[width=2.5cm]{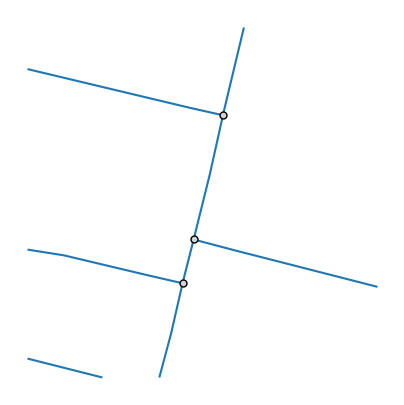}} \\
Simple intersection & Parallel edges forming an intersection. Collapse the parallel edges. & 37.06599°E 36.31055°N & \adjustbox{valign=t}{\includegraphics[width=2.5cm]{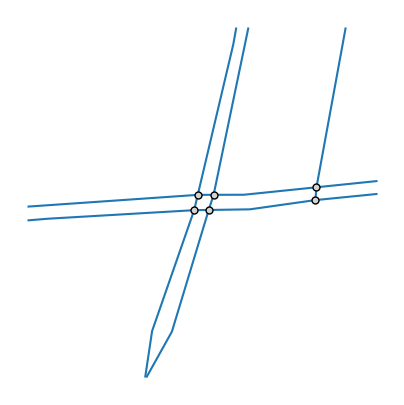}} & \adjustbox{valign=t}{\includegraphics[width=2.5cm]{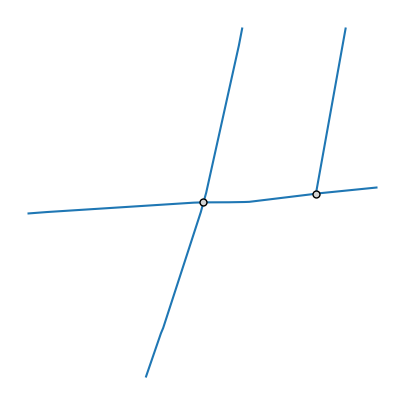}} \\
A cross-shaped intersection & Two perpendicular edges diverging before an intersection. Delete the diverging edges and snap the points of divergence to the center of the intersection. & 9.74106°E 4.09687°N & \adjustbox{valign=t}{\includegraphics[width=2.5cm]{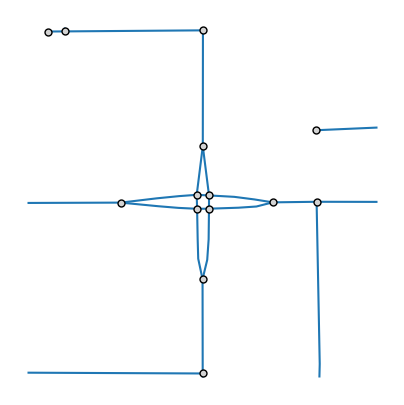}} & \adjustbox{valign=t}{\includegraphics[width=2.5cm]{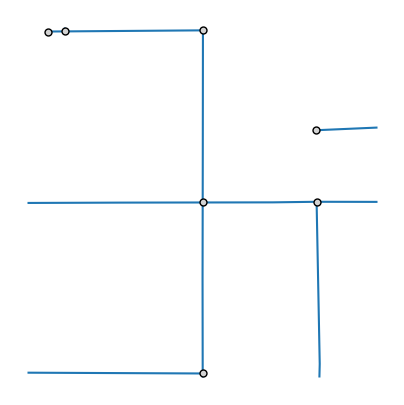}} \\
Intersection & Multiple streets meeting at the same location but not meeting at an intersecting point. Find a reasonable intersecting point (e.g.~find the centroid of the intersection or find a suitable vertex somewhere along the main street) and connect the streets. & 37.14270°E 36.23658°N & \adjustbox{valign=t}{\includegraphics[width=2.5cm]{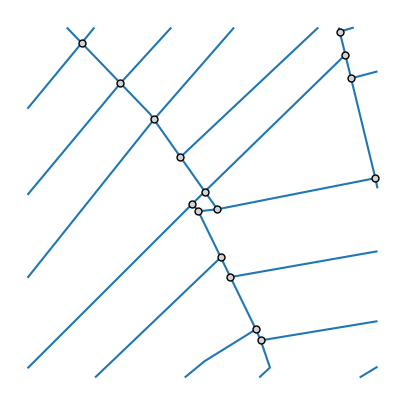}} & \adjustbox{valign=t}{\includegraphics[width=2.5cm]{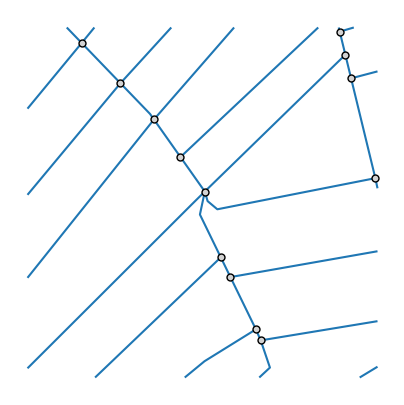}} \\
Side edges & Side edges diverging and converging to the same street. Delete if not considered relevant (e.g.~not passing through a gas station). & 37.13986°E 36.24012°N &\adjustbox{valign=t}{\includegraphics[width=2.5cm]{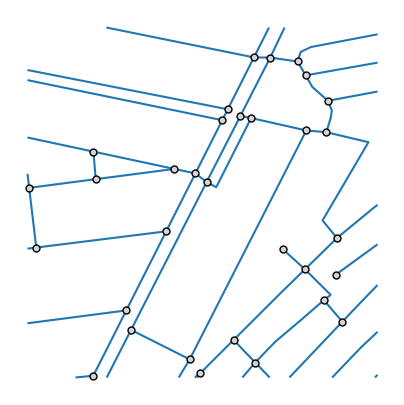}} & \adjustbox{valign=t}{\includegraphics[width=2.5cm]{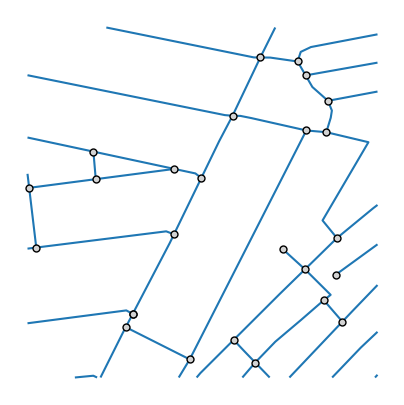}} \\
Cul-de-sac & Dead-end streets with a circular end. Snap the street to the most distant vertex of the circular part, then delete the circle. & 5.631484°E 50.605893°N & \adjustbox{valign=t}{\includegraphics[width=2.5cm]{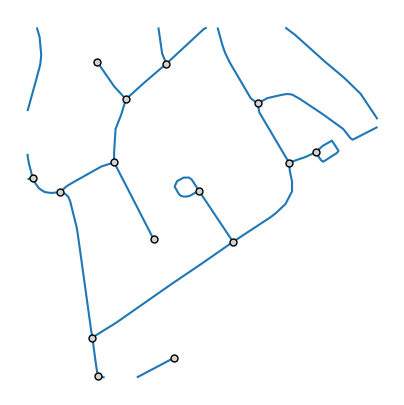}} & \adjustbox{valign=t}{\includegraphics[width=2.5cm]{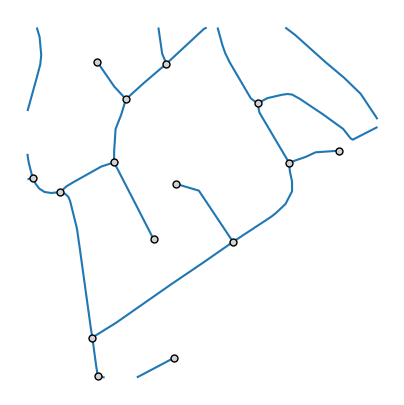}} \\
Ovalabout & Delete the ovalabout and create an intersection. The streets will intersect in the center of the original ovalabout. & 37.16807°E 36.19395°N & \adjustbox{valign=t}{\includegraphics[width=2.5cm]{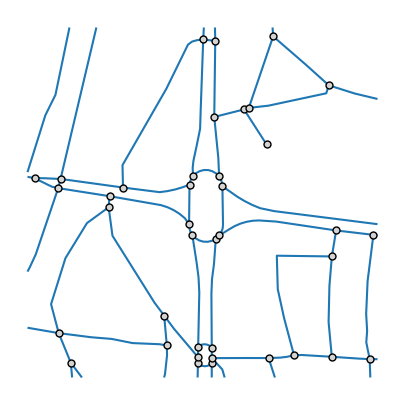}} & \adjustbox{valign=t}{\includegraphics[width=2.5cm]{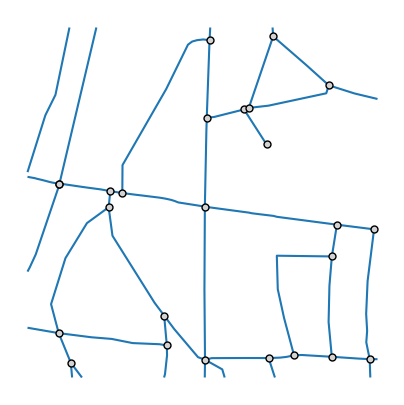}} \\
Cloverleaf interchange & Can be simplified to a basic cross-intersection if one can get from all directions to all directions. & 37.22222°E 36.19417°N & \adjustbox{valign=t}{\includegraphics[width=2.5cm]{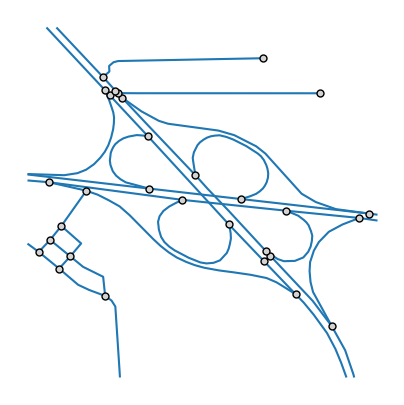}} & \adjustbox{valign=t}{\includegraphics[width=2.5cm]{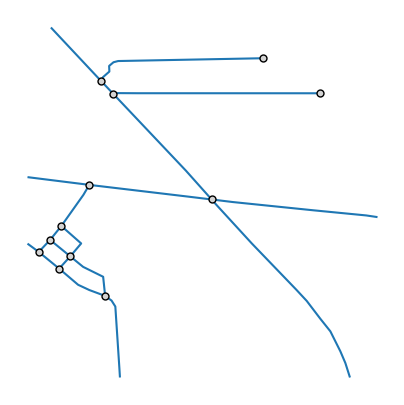}} \\
Multi-level carriageway & The intersecting streets are on different levels and therefore cannot be collapsed to a cross-intersection. & 174.84000°E 36.91861°S & \adjustbox{valign=t}{\includegraphics[width=2.5cm]{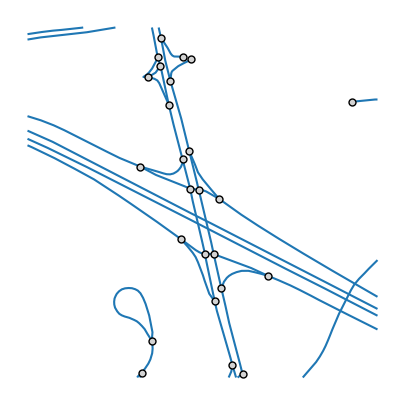}} & \adjustbox{valign=t}{\includegraphics[width=2.5cm]{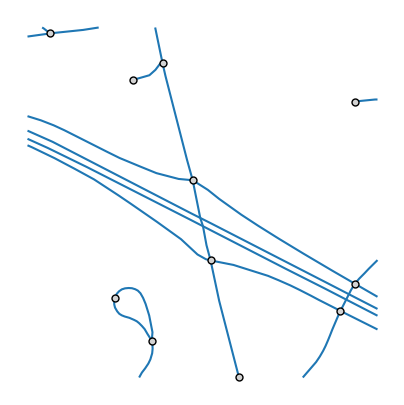}} \\
Special case roundabouts & If there is a building or a large object inside a roundabout/ovalabout/intersection, the object cannot be collapsed. & 5.61551°E 50.67645°N & \adjustbox{valign=t}{\includegraphics[width=2.5cm]{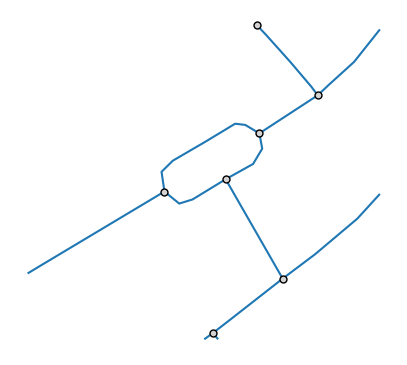}} & No change suggested \\
Parallel edges connected with a linking edge & Delete the linking edges and collapse the parallel edges. & 73.16139°W 7.06444°N & \adjustbox{valign=t}{\includegraphics[width=2.5cm]{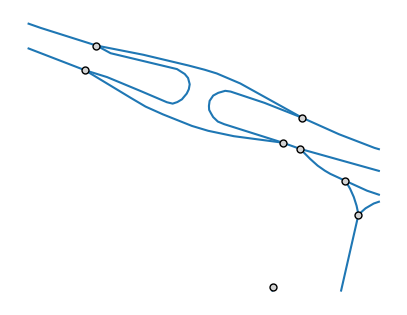}} & \adjustbox{valign=t}{\includegraphics[width=2.5cm]{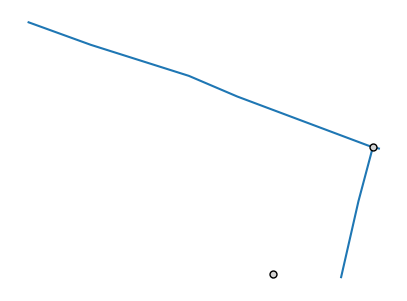}} \\
Outliers & Delete any incorrect inputs. & 73.12750°W 7.12528°N & \adjustbox{valign=t}{\includegraphics[width=2.5cm]{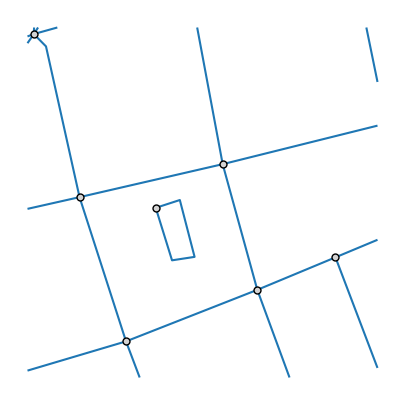}} & \adjustbox{valign=t}{\includegraphics[width=2.5cm]{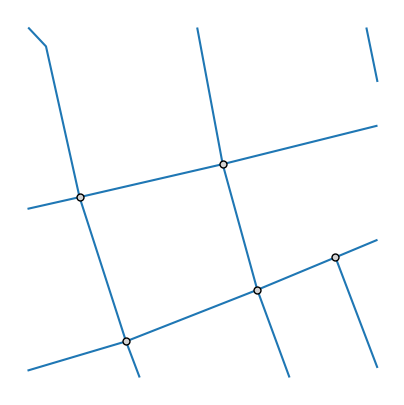}} \\
Parallel edges leading to different levels & If there are multiple parallel edges, collapse the ones on the same level, but keep the ones that are on a different level. & 9.66117°E 4.08809°N & \adjustbox{valign=t}{\includegraphics[width=2.5cm]{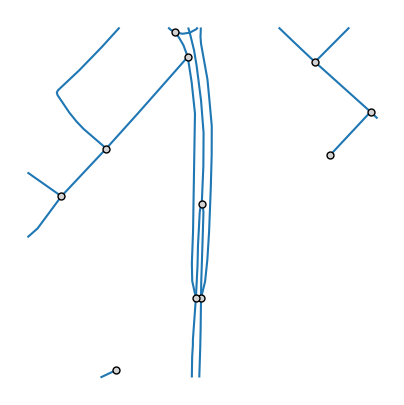}} & \adjustbox{valign=t}{\includegraphics[width=2.5cm]{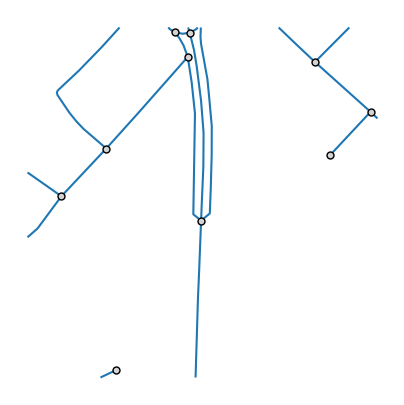}} \\
Roundabout with edges on different levels & One cannot get from one direction to all other directions, therefore it cannot be collapsed to an intersection. & 37.17978°E 36.20862°N & \adjustbox{valign=t}{\includegraphics[width=2.5cm]{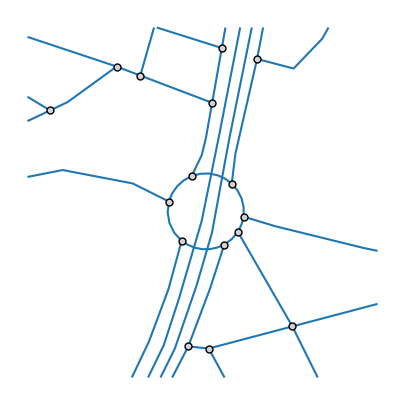}} & \adjustbox{valign=t}{\includegraphics[width=2.5cm]{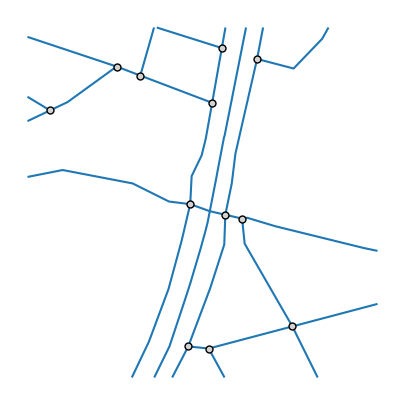}} \\
Partial cloverleaf interchange & Can be simplified to a T-intersection if one can get from all directions to all directions. & 36.99121°E 36.09677°N & \adjustbox{valign=t}{\includegraphics[width=2.5cm]{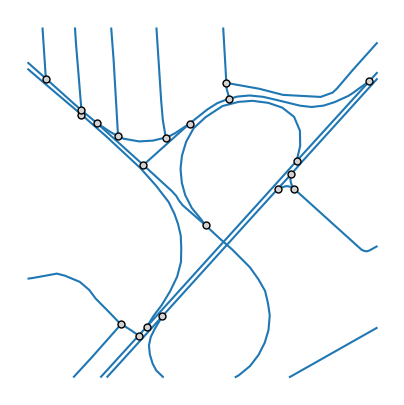}} & \adjustbox{valign=t}{\includegraphics[width=2.5cm]{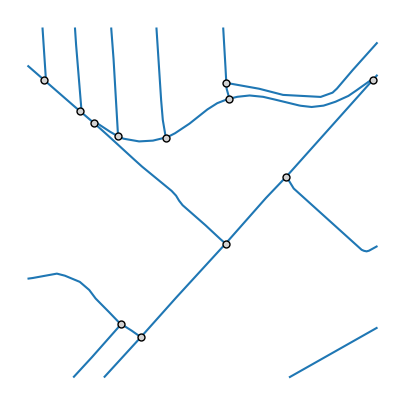}} \\
Complicated freeway intersection & There is no unified solution. & 174.7616667°E 36.7991667°S & \adjustbox{valign=t}{\includegraphics[width=2.5cm]{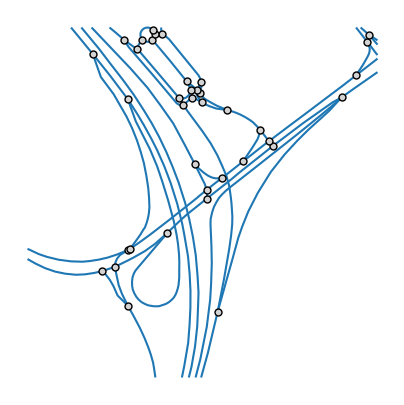}} & \adjustbox{valign=t}{\includegraphics[width=2.5cm]{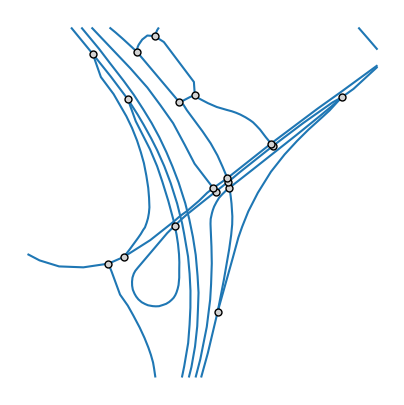}} \\
\label{tab:situations_and_descriptions}
\end{longtable}

\clearpage

\section{The \texttt{cityseer} simplification routine}\label{appendix:cityseer}

\texttt{cityseer} employs a methodical 15-step simplification routine that generally follows an iterative ``remove-smooth-reconnect'' flow. The enumerated steps are detailed in Table~\ref{tab:cityseer-routine} with a description of the operation, line number, and relevant function. This enumeration is based on \texttt{cityseer==v4.17.2}\footnote{https://github.com/benchmark-urbanism/cityseer-api/tree/v4.17.2} found on GitHub in \texttt{io.py\#L250}\footnote{https://github.com/benchmark-urbanism/cityseer-api/blob/v4.17.2/pysrc/cityseer/tools/io.py\#L250}. Further details can be found in Section 5.1 of \citet{simons_cityseer_2022} and the \texttt{cityseer} Graph Cleaning Guide\footnote{https://benchmark-urbanism.github.io/cityseer-examples/examples/graph\_cleaning.html\#manual-cleaning}.

\begin{table}[h!]
\centering
\begin{tabular}{rlrl}
step & description & line number & relevant function\\
\midrule
1  & initial deduplication                                                                  & 260       & \texttt{graphs.nx\_deduplicate\_edges}        \\
2  & exclusions and removals                                                                & 262--329  & --                                            \\
3  & removal of dangling nodes                                                              & 331       & \texttt{graphs.nx\_remove\_dangling\_nodes}   \\
4  & \makecell{splitting opposing geometries\\\qquad(iterated over 4 sets of parameters)}   & 340       & \texttt{graphs.nx\_split\_opposing\_geoms}    \\
5  & \makecell{consolidate nodes\\\qquad(iterated over 4 sets of parameters)}               & 357       & \texttt{graphs.nx\_consolidate\_nodes}        \\
6  & removal of degree-2 nodes                                                              & 367       & \texttt{graphs.nx\_remove\_filler\_nodes}     \\
7  & snap gapped endings                                                                    & 369       & \texttt{graphs.nx\_snap\_gapped\_endings}     \\
8  & splitting opposing geometries                                                          & 388       & \texttt{graphs.nx\_split\_opposing\_geoms}    \\
9  & removal of dangling nodes                                                              & 416       & \texttt{graphs.nx\_remove\_dangling\_nodes}   \\
10 & \makecell{splitting opposing geometries\\\qquad(iterated over twice)}                  & 420       & \texttt{graphs.nx\_split\_opposing\_geoms}    \\
11 & \makecell{consolidate nodes\\\qquad(iterated over twice)}                              & 448       & \texttt{graphs.nx\_consolidate\_nodes}        \\
12 & removal of degree-2 nodes                                                              & 477       & \texttt{graphs.nx\_remove\_filler\_nodes}     \\
13 & merging parallel edges                                                                 & 478       & \texttt{graphs.nx\_merge\_parallel\_edges}    \\
14 & ironing out edges                                                                      & 479       & \texttt{graphs.nx\_iron\_edges}               \\
15 & removal of dangling nodes                                                              & 481       & \texttt{graphs.nx\_remove\_dangling\_nodes}   \\
\bottomrule
\end{tabular}
\caption{The \texttt{cityseer} simplification routine described as found in \texttt{cityseer/tools/io.py\#L250} of \texttt{v4.17.2}.}
\label{tab:cityseer-routine}
\end{table}

\clearpage

\section{CES typology of Liège}\label{appendix:ces}

\begin{figure}[h]
    \centering
    \includegraphics[width=0.95\textwidth]{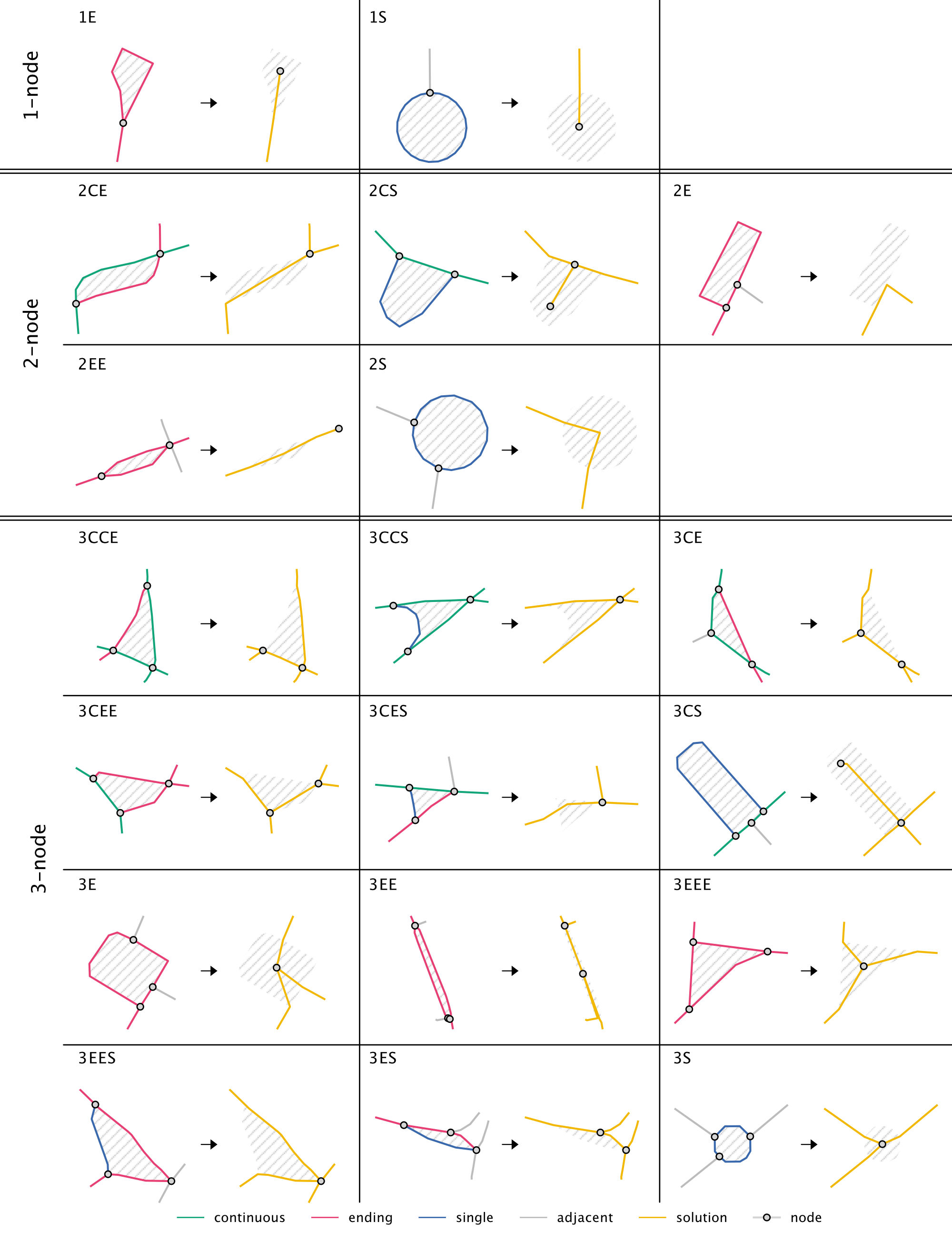}
    \caption{CES typology derived from Liège street network (part 1).}
    \label{fig:ces1}
\end{figure}

\begin{figure}[h]
    \centering
    \includegraphics[width=0.95\textwidth]{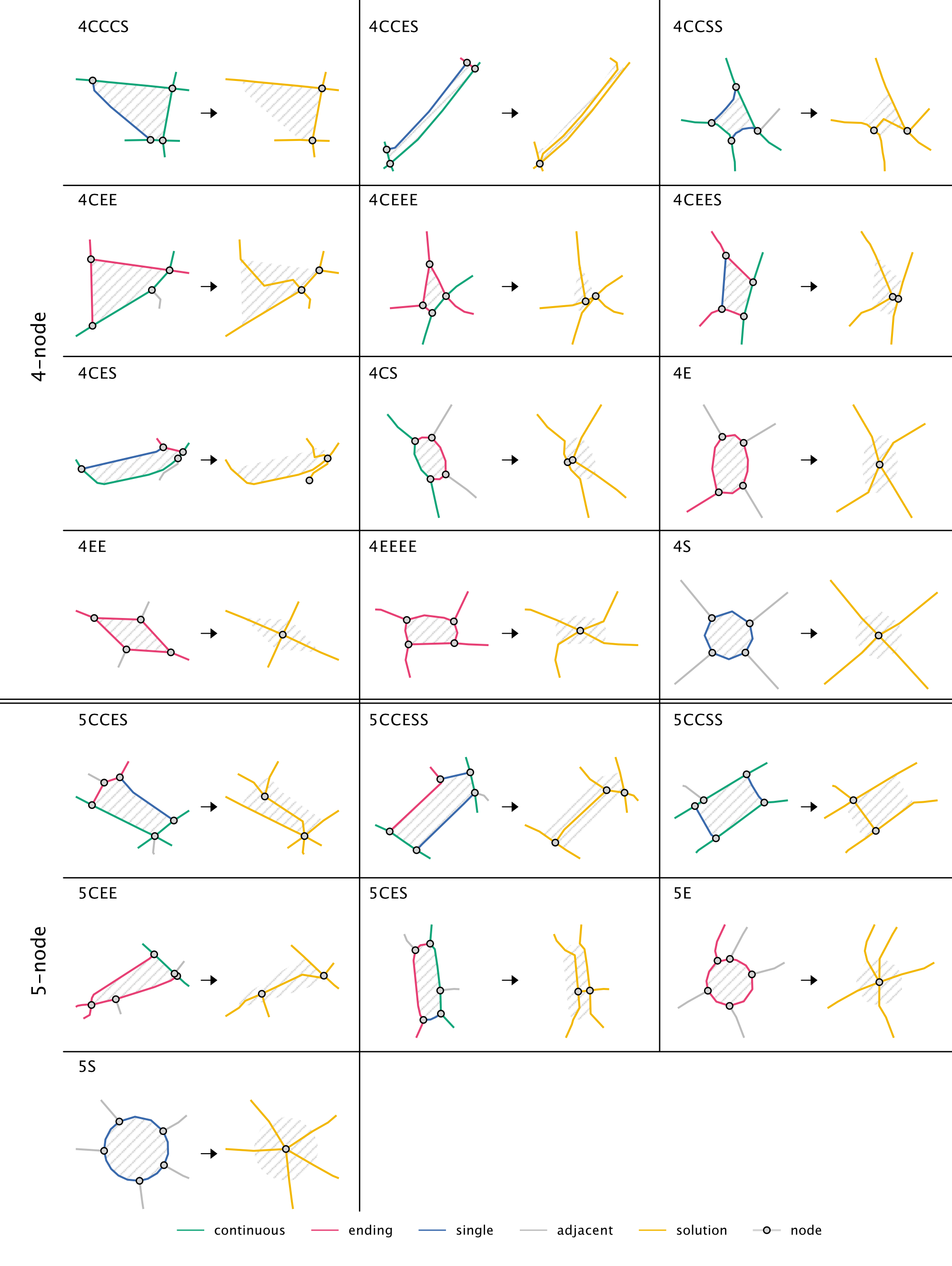}
    \caption{CES typology derived from Liège street network (part 2).}
    \label{fig:ces2}
\end{figure}

\begin{figure}[h]
    \centering
    \includegraphics[width=0.95\textwidth]{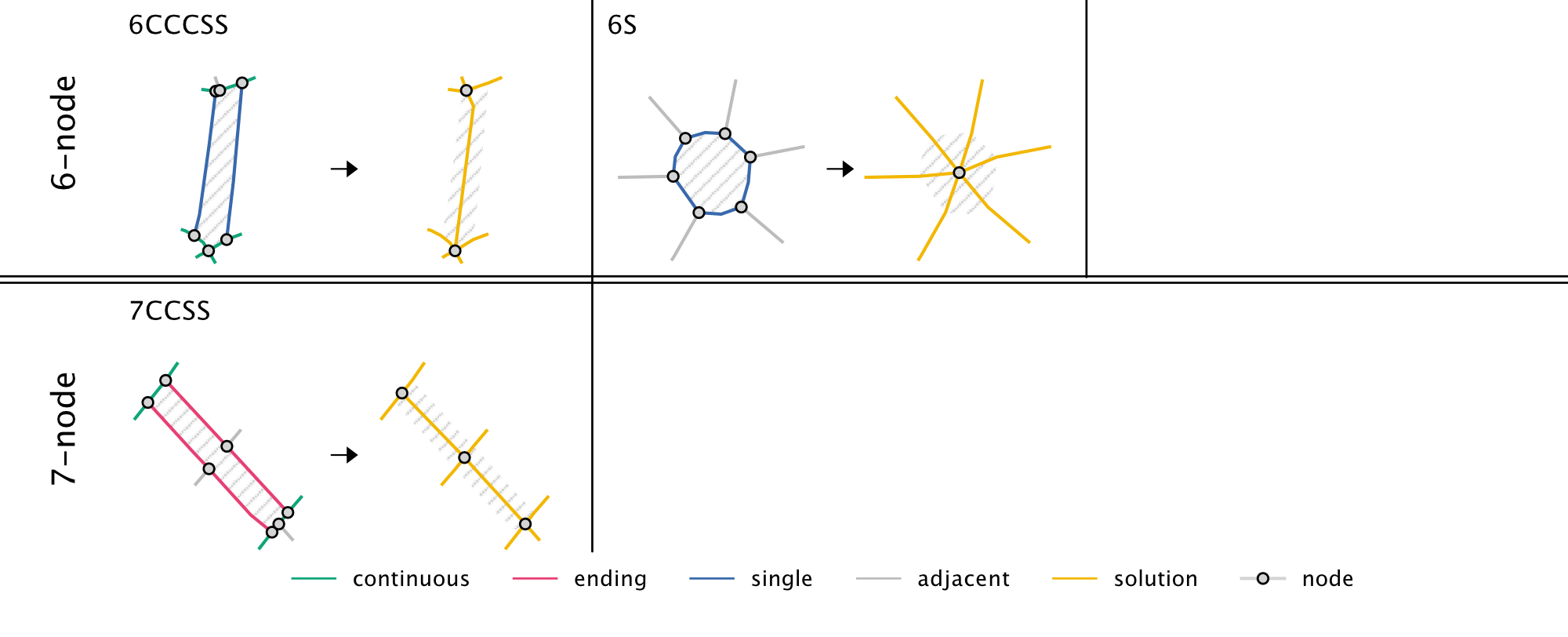}
    \caption{CES typology derived from Liège street network (part 3).}
    \label{fig:ces3}
\end{figure}

Figures \ref{fig:ces1}-\ref{fig:ces3} show the complete list of CES types observed in the Liège study area with a single example of each type and its proposed solution. A single CES type is composed of the number of nodes forming the artifact and the continuity types of strokes forming its boundary. For example, the most common type, 3CES, is composed of 3 nodes and 3 continuity strokes (of types C, E, and S), while 5S is composed of 5 nodes and a single continuity stroke of type S (as it never leaves the artifact). The total number of CES types is theoretically infinite, given no upper bound on the number of nodes exists.

\clearpage

\section{Use case solutions by simplification method}\label{appendix:solutions}

In Table~\ref{tab:solutions} below, for the 19 use cases from the manual simplification protocol in Appendix~\ref{appendix:protocol}, we plot the input data for each case (first column), and the solutions to each case by method (evaluated methods: \texttt{cityseer}, \texttt{OSMnx}, and \texttt{parenx}; our proposed method \texttt{neatnet}; and finally, the manually simplified case).

\begin{longtable}{c}
\caption{Solutions by method (columns) for each of the use cases (rows) from the manual simplification protocol in Appendix~\ref{appendix:protocol}.} \\
\hline
\adjustbox{valign=t}{\includegraphics[width=\linewidth]{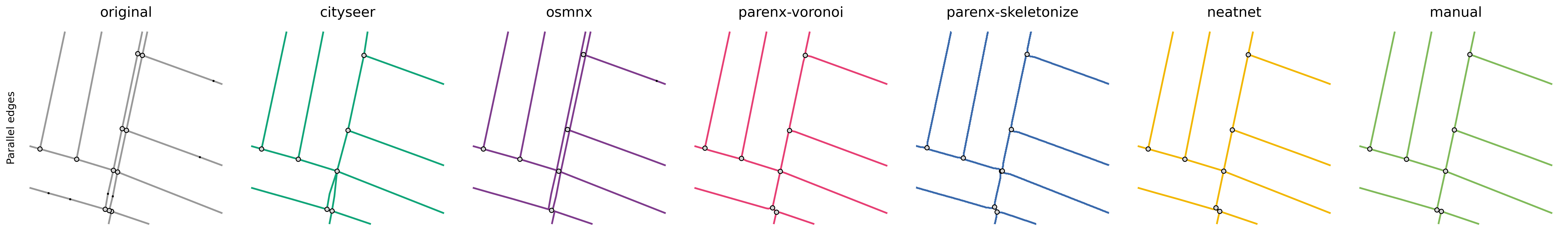}} \\
\hline
\adjustbox{valign=t}{\includegraphics[width=\linewidth]{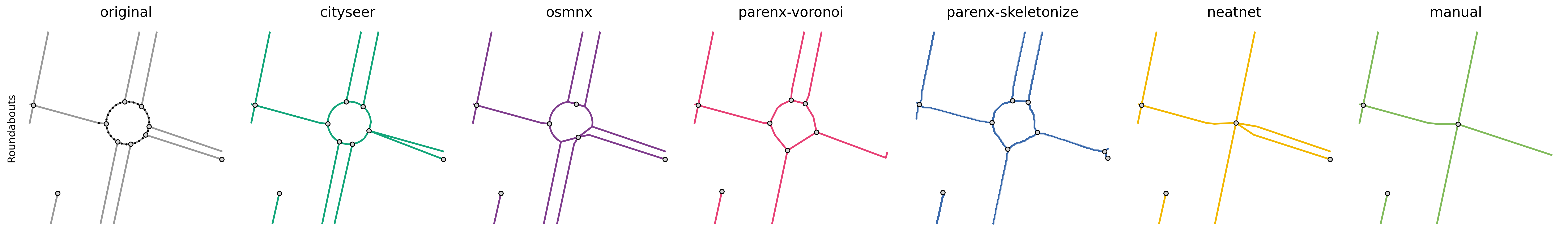}} \\
\hline
\adjustbox{valign=t}{\includegraphics[width=\linewidth]{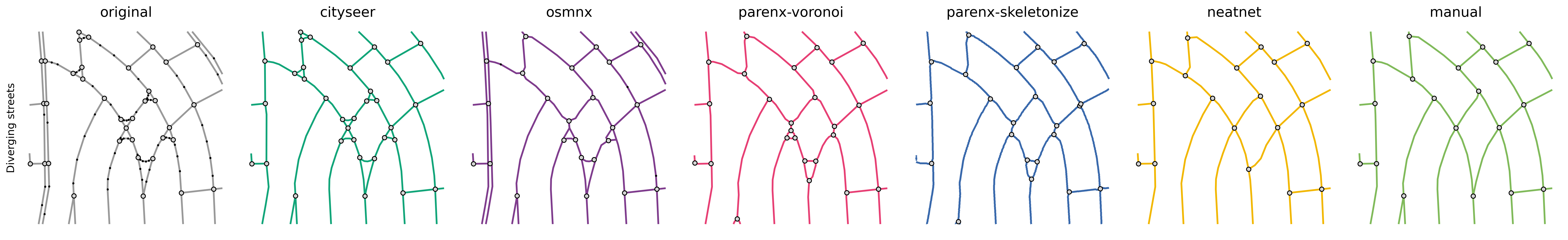}} \\
\hline
\adjustbox{valign=t}{\includegraphics[width=\linewidth]{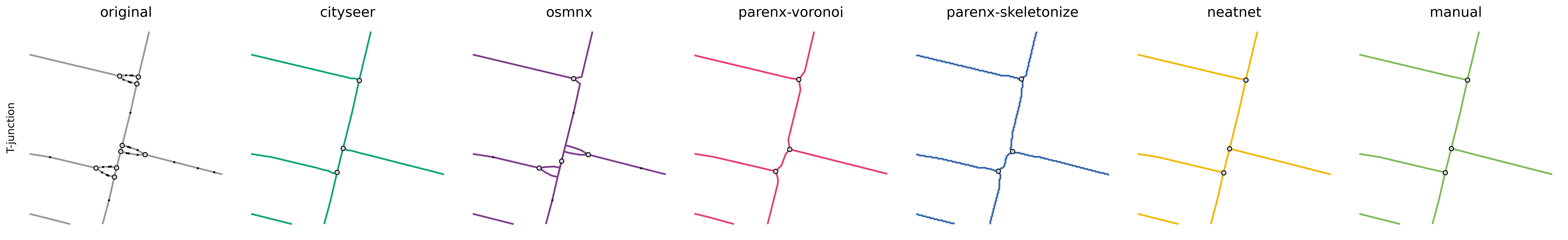}} \\
\hline
\adjustbox{valign=t}{\includegraphics[width=\linewidth]{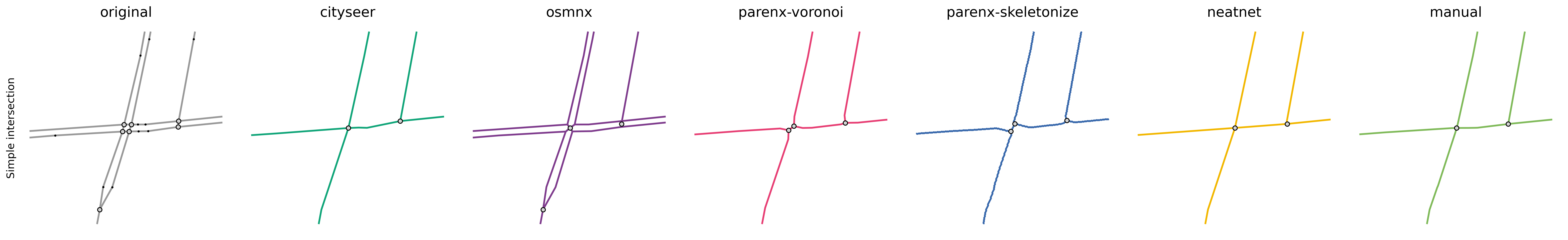}} \\
\hline
\adjustbox{valign=t}{\includegraphics[width=\linewidth]{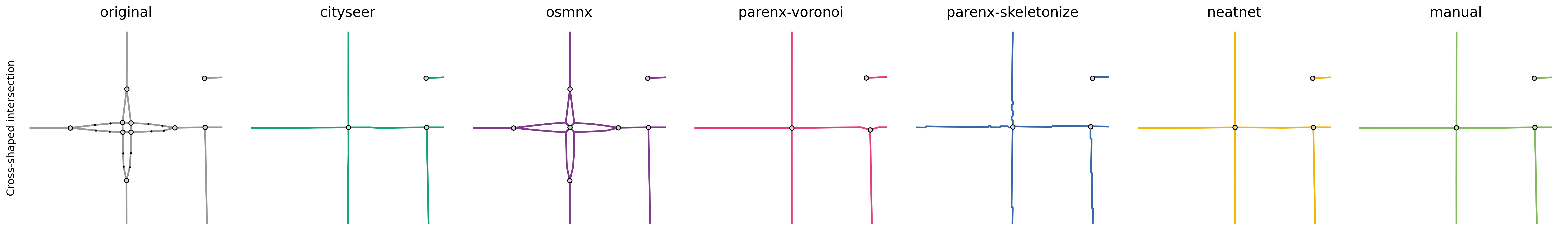}} \\
\hline
\adjustbox{valign=t}{\includegraphics[width=\linewidth]{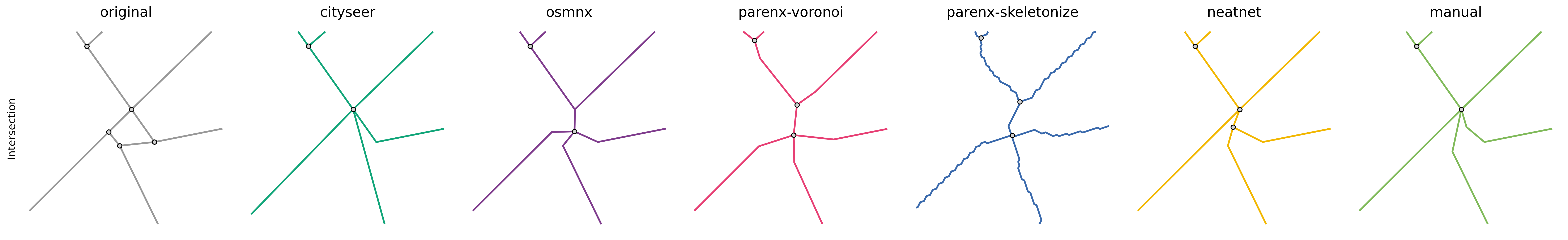}} \\
\hline
\adjustbox{valign=t}{\includegraphics[width=\linewidth]{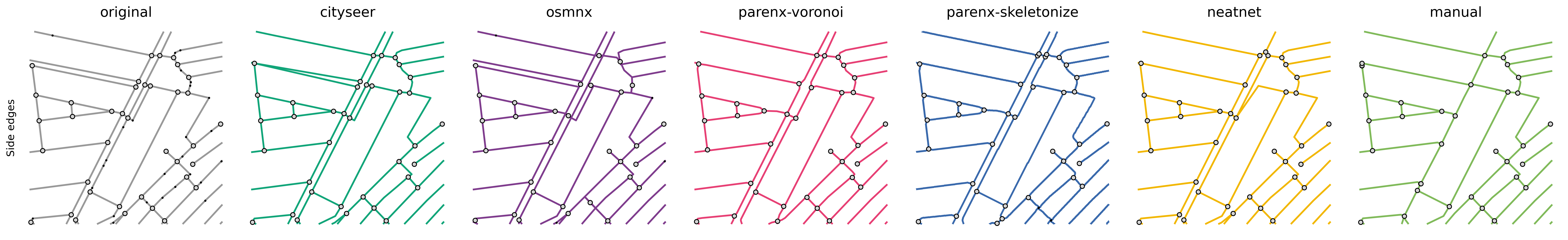}} \\
\hline
\adjustbox{valign=t}{\includegraphics[width=\linewidth]{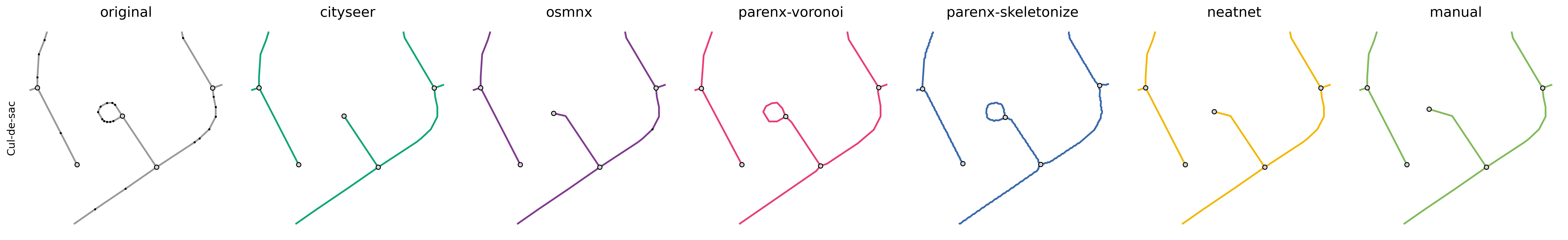}} \\
\hline
\adjustbox{valign=t}{\includegraphics[width=\linewidth]{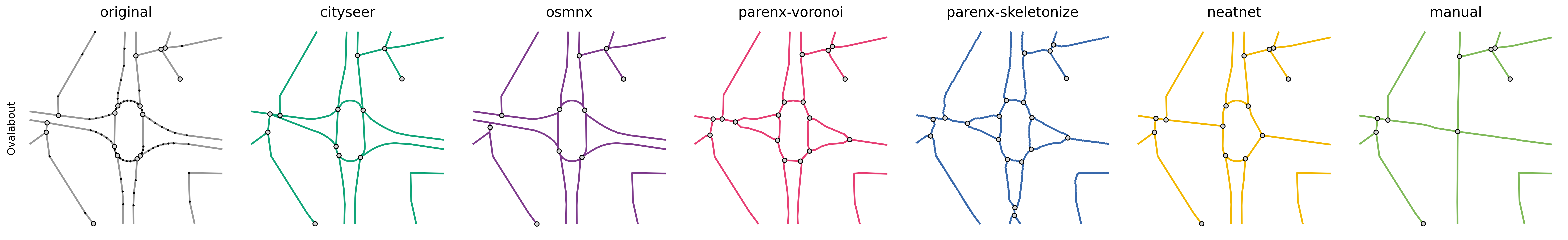}} \\
\hline
\adjustbox{valign=t}{\includegraphics[width=\linewidth]{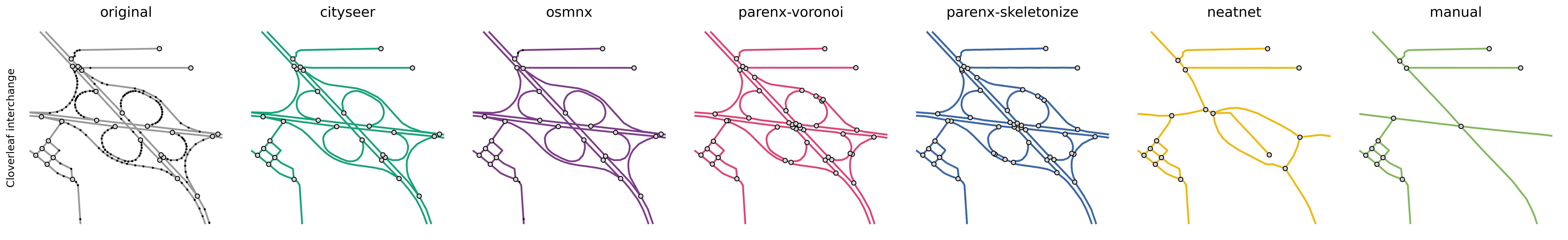}} \\
\hline
\adjustbox{valign=t}{\includegraphics[width=\linewidth]{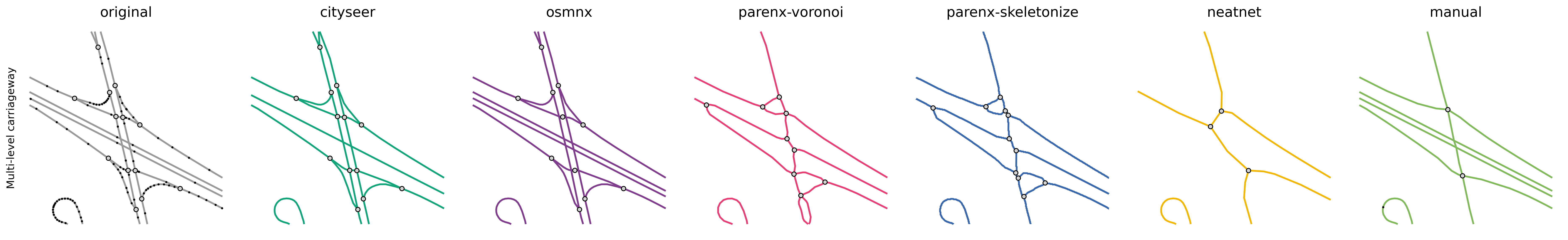}} \\
\hline
\adjustbox{valign=t}{\includegraphics[width=\linewidth]{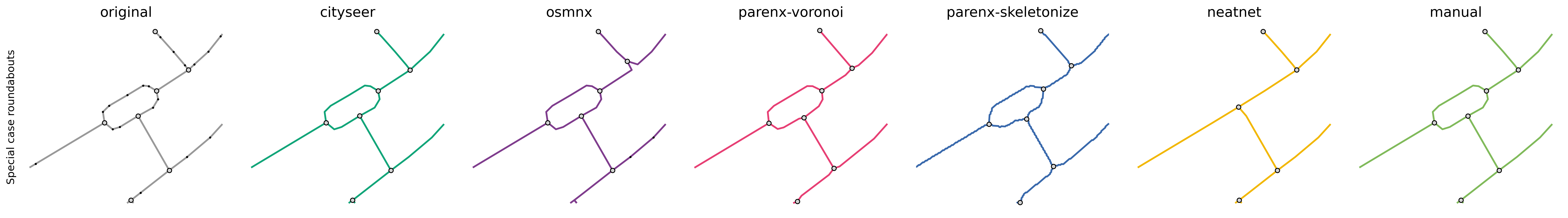}} \\
\hline
\adjustbox{valign=t}{\includegraphics[width=\linewidth]{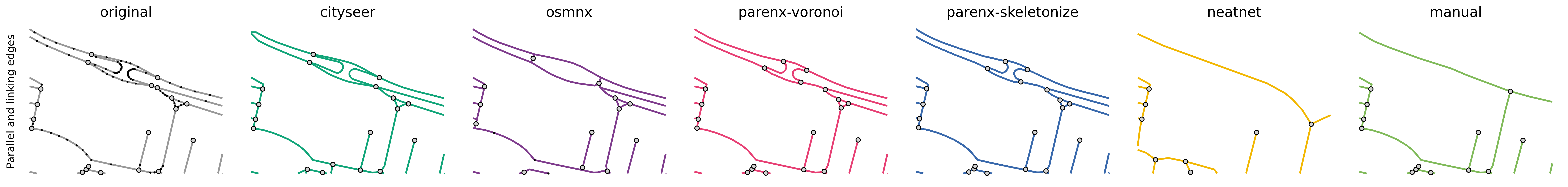}} \\
\hline
\adjustbox{valign=t}{\includegraphics[width=\linewidth]{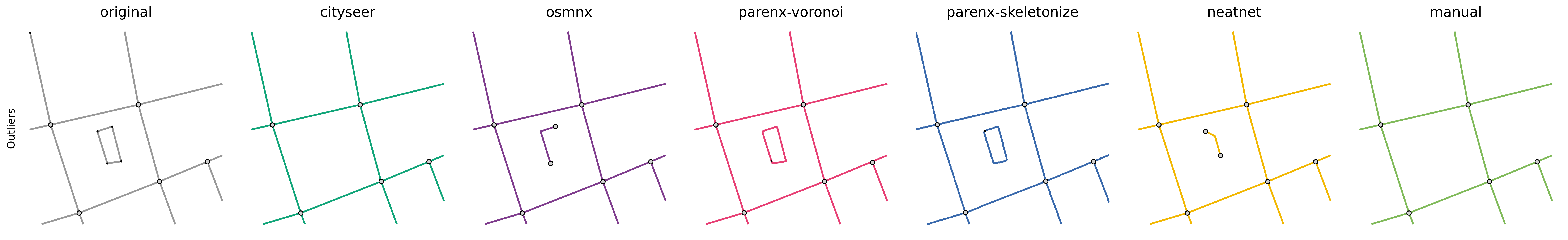}} \\
\hline
\adjustbox{valign=t}{\includegraphics[width=\linewidth]{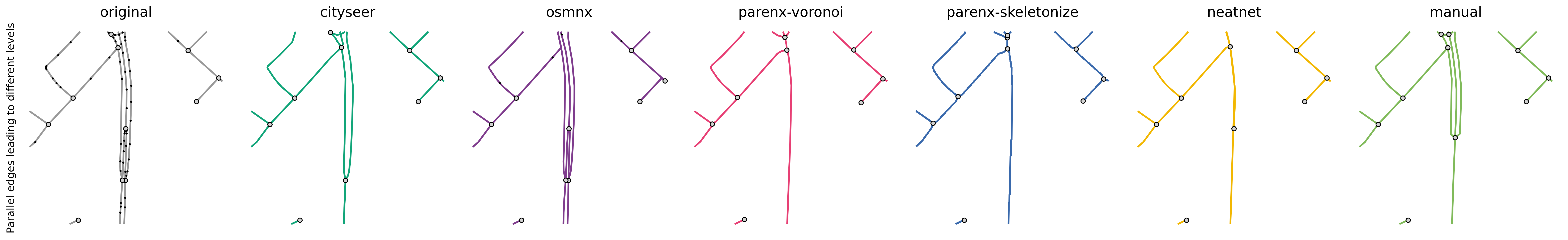}} \\
\hline
\adjustbox{valign=t}{\includegraphics[width=\linewidth]{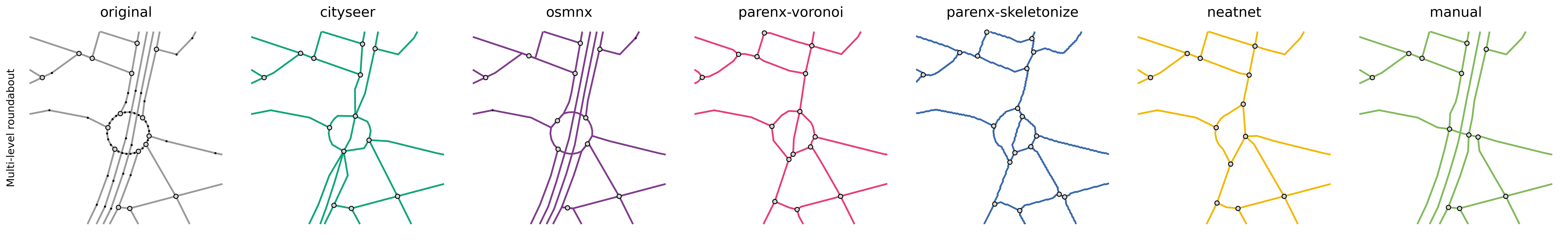}} \\
\hline
\adjustbox{valign=t}{\includegraphics[width=\linewidth]{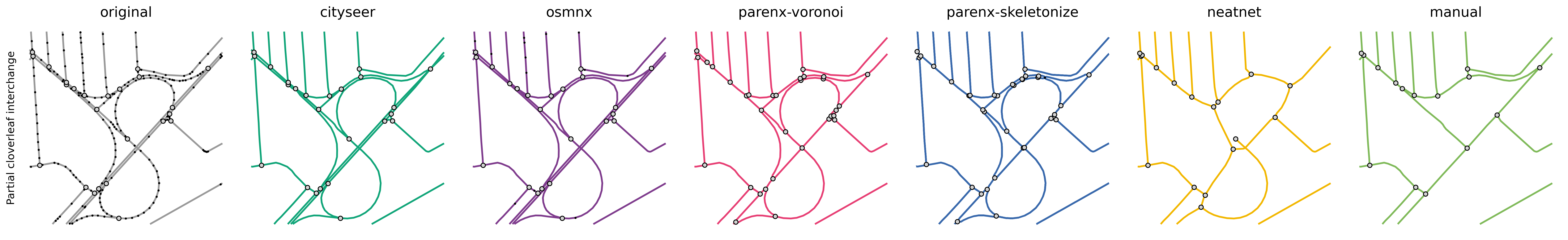}} \\
\hline
\adjustbox{valign=t}{\includegraphics[width=\linewidth]{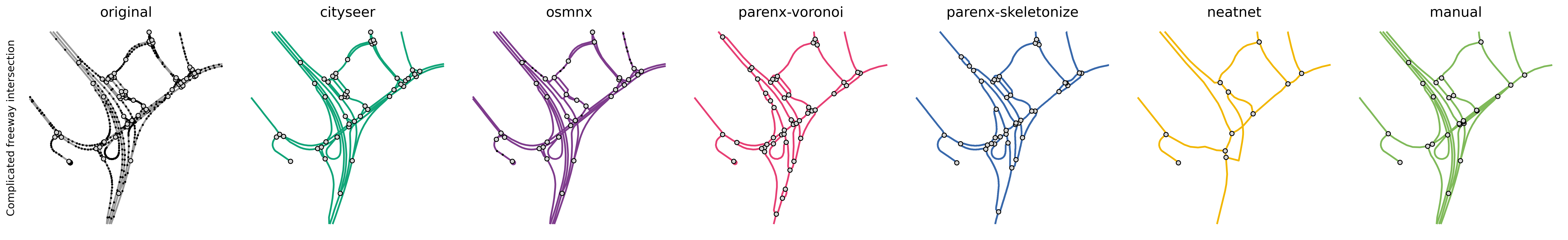}} \\

\label{tab:solutions}
\end{longtable}

\clearpage

\section{Comparative illustration of method outputs}\label{appendix:vis-inspection}

\begin{figure}[h!]
    \centering
    \includegraphics[width=0.6\textwidth]{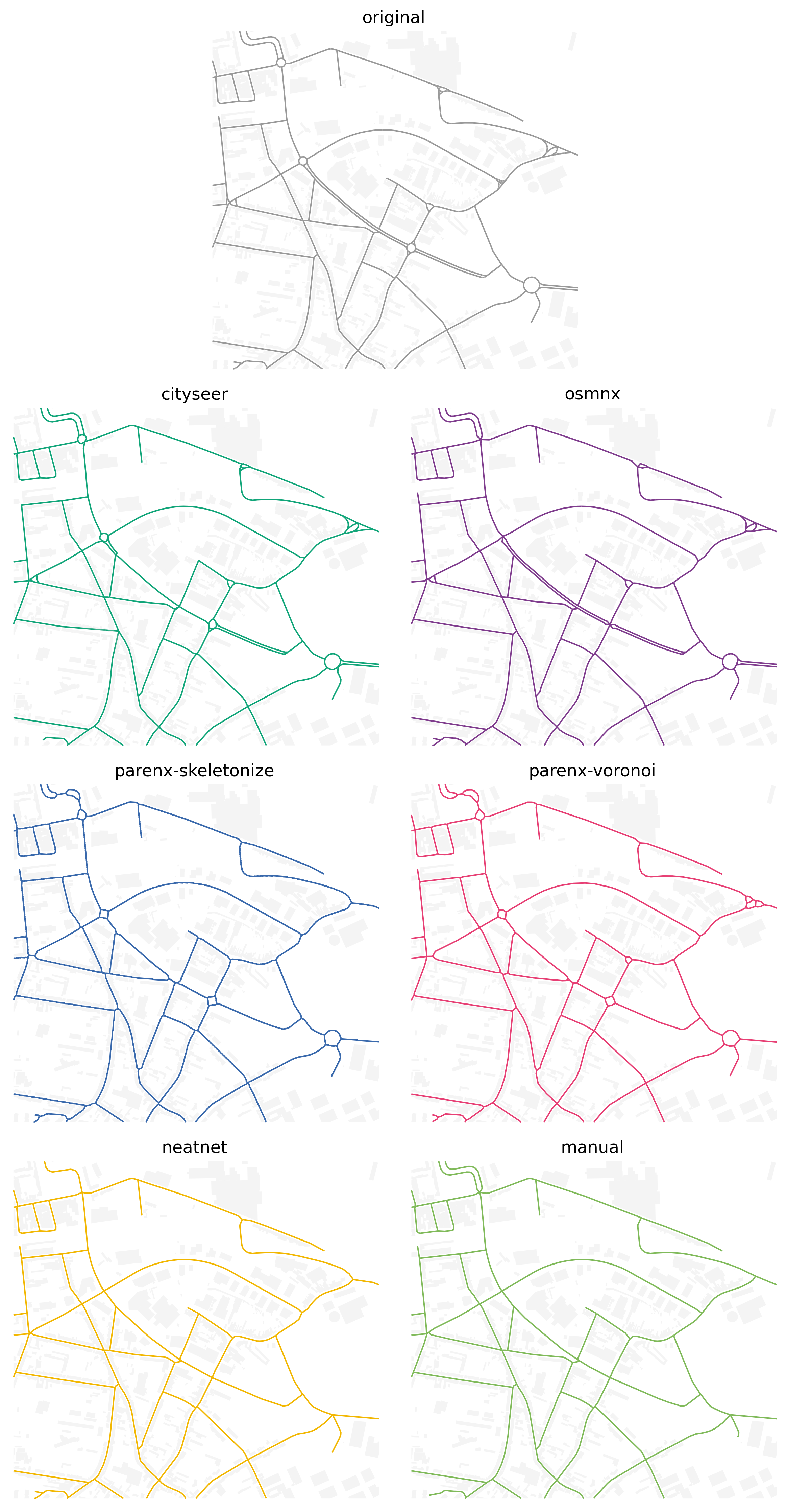}
    \caption{Output of different simplification methods for a street network fragment in Seraing, Liège (Belgium).}
    \label{fig:allmethods}
\end{figure}

For a better visual grasp of the qualitative differences between simplification methods compared in this study (\texttt{cityseer}, \texttt{OSMnx}, \texttt{parenx}, \texttt{neatnet}, and manual simplification), Figure~\ref{fig:allmethods} shows the output of each method for the example case presented in Figure~\ref{fig:original-manual} of the main text. 

\clearpage

\section{Simplification performance using Pearson and Spearman correlation}\label{appendix:corr}

The section \ref{sec:results} showcases the outcome of simplification performance evaluation using the Chatterjee's $\xi$ coefficient, which on this data proves to be more efficient in capturing the differences between the methods. Below are the results of Pearson's and Spearman's correlation coefficients for a reference.

\begin{figure}[h]
    \centering
    \includegraphics[width=0.95\textwidth]{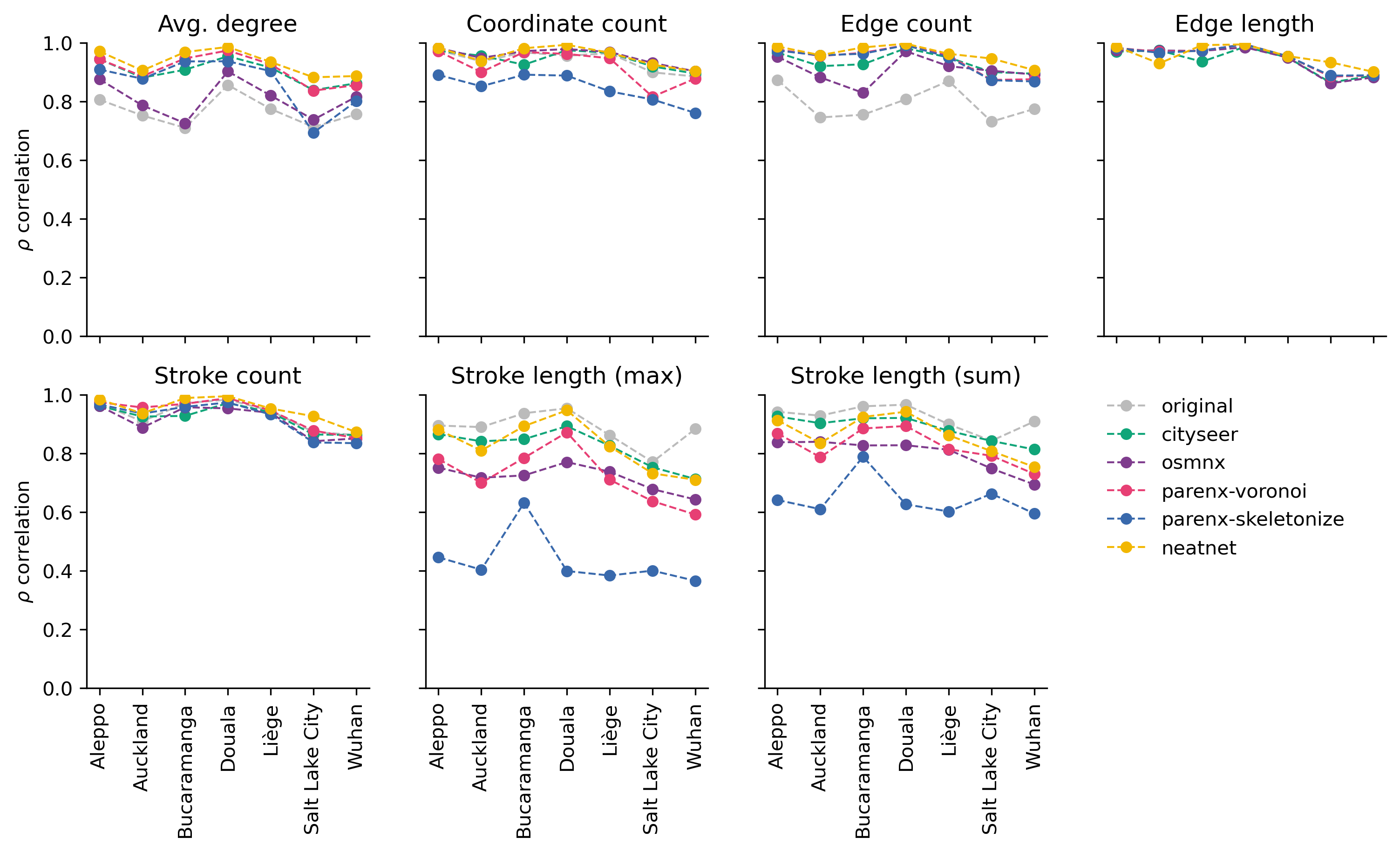}
    \caption{Pearson's $\rho$ correlation coefficient between properties of manually simplified networks and networks based on each of the tested algorithms vs.~the original network as a baseline. Higher is considered better -- approaching $1.0$ on the $y$-axis.}
    \label{fig:rho}
\end{figure}

\begin{figure}[h]
    \centering
    \includegraphics[width=0.95\textwidth]{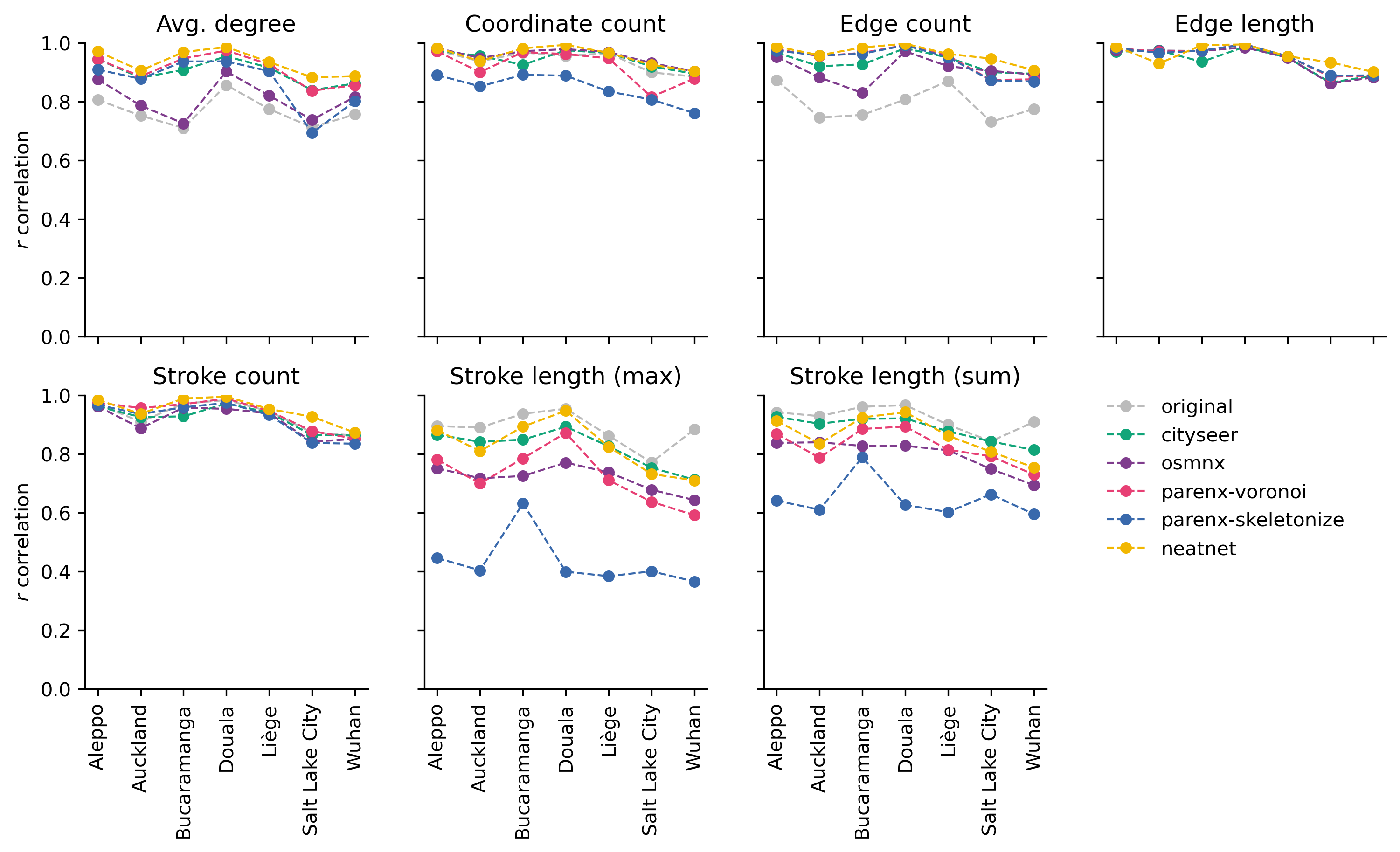}
    \caption{Spearman's rank correlation coefficient between properties of manually simplified networks and networks based on each of the tested algorithms vs.~the original network as a baseline. Higher is considered better -- approaching $1.0$ on the $y$-axis.}
    \label{fig:r}
\end{figure}

\clearpage

\section{Evaluation with building data}\label{appendix:eval-with-buildings}

\begin{figure}[h]
    \centering
    \includegraphics[width=0.95\textwidth]{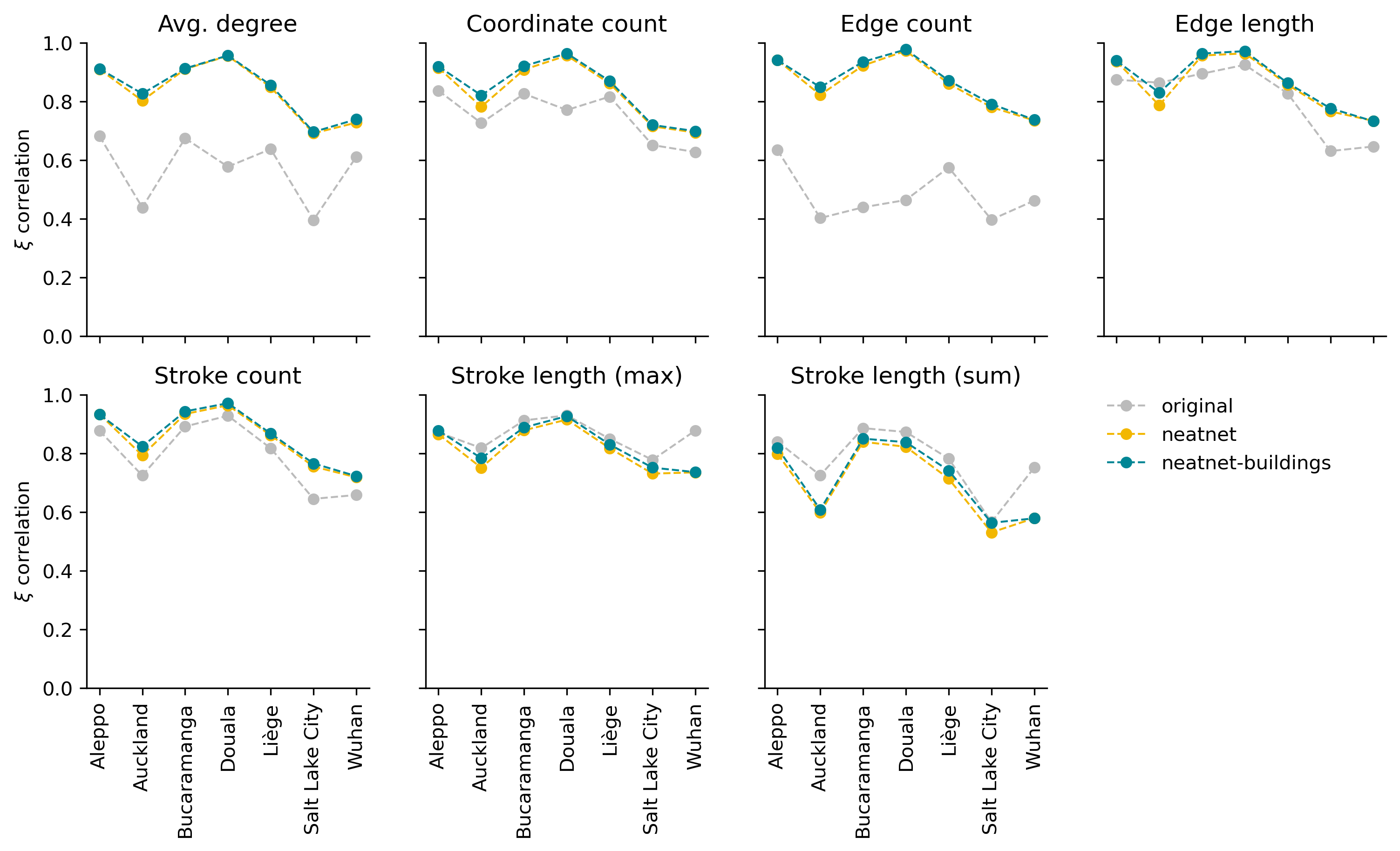}
    \caption{Chatterjee's $\xi$ correlation coefficient between properties of networks that are~(a)~manually simplified (``manual''),~(b)~simplified by our proposed method (``neatnet''),~(c)~simplified by our proposed method after applying a building exclusion mask (``neatnet-buildings''),~vs.~the original network as a baseline. Higher is considered better -- approaching $1.0$ on the $y$-axis. For all seven considered metrics and in all seven considered cities, adding the building exclusion mask slightly increases the $\xi$ correlation coefficient, bringing it closer to a value of $1.0$.
    }
    \label{fig:chi-buildings}
\end{figure}

\begin{figure}[h]
    \centering
    \includegraphics[width=0.95\textwidth]{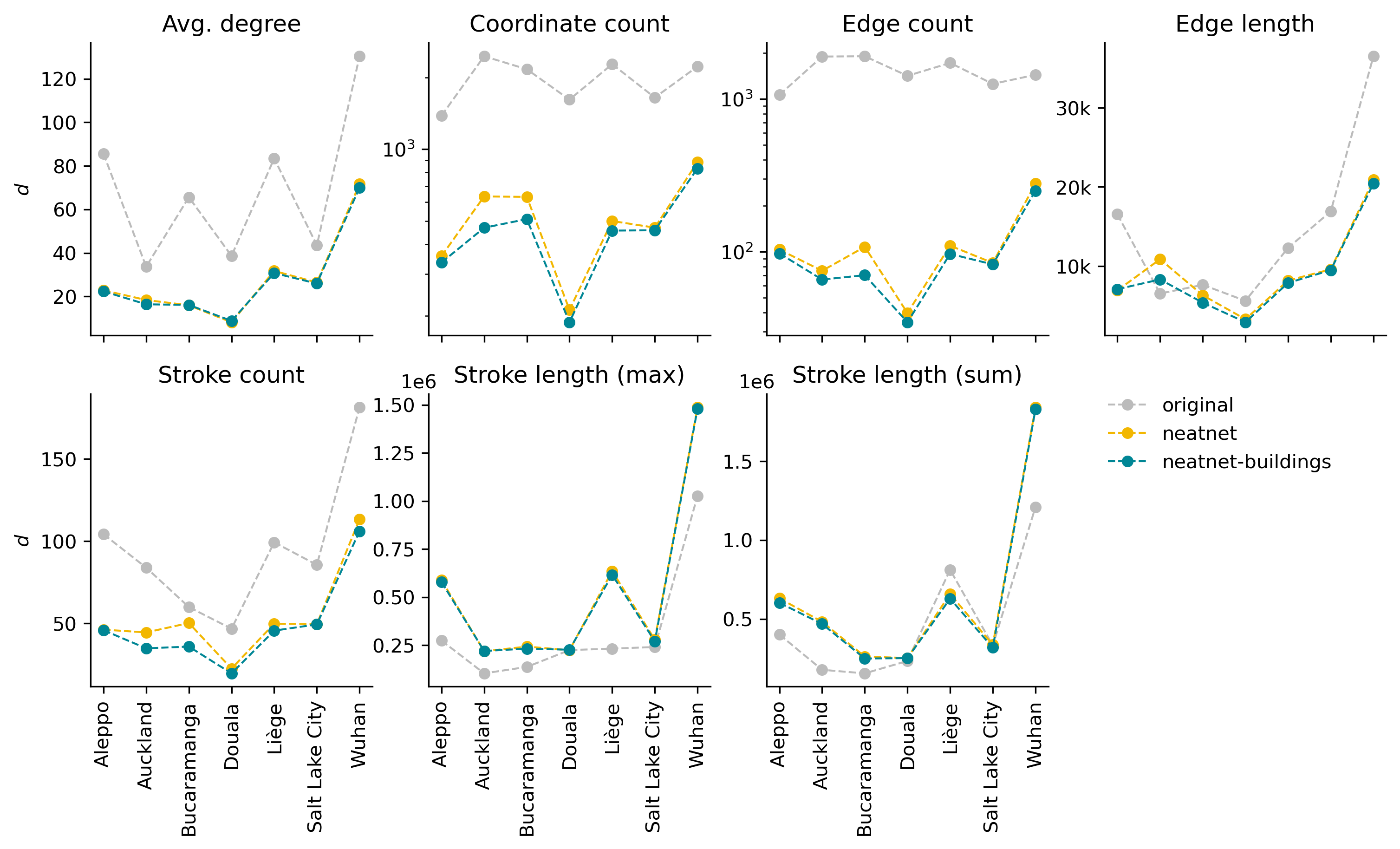}
    \caption{Euclidean distance between properties of networks that are~(a)~manually simplified (``manual''),~(b)~simplified by our proposed method (``neatnet''),~(c)~simplified by our proposed method after applying a building exclusion mask (``neatnet-buildings''), vs.~the original network as a baseline. Lower is considered better -- approaching $0.0$ on the $y$-axis. For all seven considered metrics and in all seven considered cities, adding the building exclusion mask slightly decreases the Euclidean distance between the distributions, bringing it closer to a value of $0.0$.
    }
    \label{fig:delta-buildings}
\end{figure}

The comparative evaluation of different simplification methods (see Section~\ref{sec:evaluation} in the main text) uses minimal input data (only LineString geometries representing the street network). However, if and when additional data is available for a given location, its incorporation into the simplification process can help obtain even better results. Our proposed method includes an option to use an exclusion mask, where the user can indicate which polygons representing the face artifacts of the network should be \textit{conserved} in the simplified network. Here, we use as an exclusion mask of all face artifacts that contain (intersect with) building footprint polygon data derived from OpenStreetMap \cite{openstreetmap_contributors_openstreetmap_2025}. The rationale behind this approach is that any face polygon that contains or intersects with a building footprint is more likely to be a proper urban block than a face artifact~\cite{fleischmann_shape-based_2024}, and its delineating edges should therefore \texttt{not} be changed by the simplification process. Comparing the quality of results of our proposed method with vs.~without the exclusion mask for buildings simultaneously provides an estimate of the false positive rate of our method for the latter case. Figures~\ref{fig:chi-buildings} and~\ref{fig:delta-buildings} compare the $\xi$ correlation coefficient and the absolute deviation, respectively, for our method with vs.~without building data, showing that using a building exclusion mask slightly improves the final outcome of the simplification process.

\end{document}